\documentclass{raa_rb}

\newcommand{\HII}{$\rm H~{\scriptstyle II}$}
\newcommand{\HI}{$\rm H~{\scriptstyle I}$}
\newcommand{\MgI}{$\rm Mg~{\scriptstyle I}$}
\newcommand{\kms}{$\rm km\,s^{-1}$}
\newcommand{\bsnr}{$\rm SNR(4750\,\AA)$}
\newcommand{\rsnr}{$\rm SNR(7450\,\AA)$}

\usepackage[percent]{overpic}
\usepackage{times}
\usepackage{graphicx}
\usepackage{amsmath}
\usepackage{mathabx}
\usepackage{subfigure}
\usepackage{natbib}
\usepackage{txfonts}
\usepackage{journals}
\usepackage{color}
\usepackage{colordvi}
\usepackage{ulem}

\begin{document}

\title
{The LAMOST Spectroscopic Survey of Globular Clusters in M\,31 and M\,33.
I. Catalog and new identifications}

   \volnopage{Vol.0 (200x) No.0, 000--000}      
   \setcounter{page}{1}          

   \author{B.-Q. Chen
      \inst{1}\thanks{LAMOST Fellow.}
   \and X.-W. Liu
      \inst{1,2}
   \and  M.-S. Xiang
      \inst{1}
   \and   H.-B. Yuan
      \inst{2}\footnotemark[1]
   \and  Y. Huang
      \inst{1}
   \and Z.-Y Huo
      \inst{3}
   \and  N.-C. Sun
      \inst{1}
   \and  C. Wang
      \inst{1}
   \and  J.-J. Ren
      \inst{1}\footnotemark[1]
   \and  H.-W Zhang
      \inst{1}
   \and A. Rebassa-Mansergas
      \inst{2}\footnotemark[1]
   \and M. Yang
      \inst{4}\footnotemark[1]
   \and  Y. Zhang
      \inst{5}
   \and  Y.-H. Hou
      \inst{5}
   \and  Y.-F. Wang
      \inst{5}
   }

\institute{Department of Astronomy, Peking University, Beijing 100871, P.\,R.\,China; {\it bchen@pku.edu.cn}, {\it x.liu@pku.edu.cn}
\\
  \and
Kavli Institute for Astronomy and Astrophysics, Peking University, Beijing 100871, P.\,R.\,China
\\
  \and
  National Astronomical Observatories, Chinese Academy of Sciences, Beijing 100012, P.\,R.\,China
\\
\and
 Key Laboratory of Optical Astronomy, National Astronomical Observatories, Chinese Academy of Sciences, Beijing 100012, China
\\
\and
 Nanjing Institute of Astronomical Optics \& Technology, National Astronomical Observatories, Chinese Academy of Sciences, Nanjing 210042, China
   }

   \date{Received~~2015 month day; accepted~~2015~~month day}

\abstract{
We present a catalog of {908} objects observed with the Large Sky Area Multi-Object
Fiber Spectroscopic Telescope (LAMOST)  in the vicinity fields of M31 and M33, 
targeted as globular clusters (GCs) and candidates. 
The targets include known GCs and candidates selected from the literature, as well as  
new candidates selected from  the Sloan Digital Sky Survey (SDSS). 
Analysis shows that {356} of them are likely  GCs of various degree of confidence, 
while the remaining ones turn out to be background galaxies and quasars, 
stars {and} \HII~ regions in M31 or foreground {Galactic} stars.
The {356} likely GCs include 298 {bona fide} GCs
and 26 candidates known in the literature. 
Three candidates selected from the Revised Bologna Catalog of M31 GCs and candidates (RBC) and 
one possible cluster from Johnson et al. are confirmed to be bona fide clusters. 
We search for new GCs in the halo of the M31 amongst the new candidates 
 selected {from} the  SDSS photometry. 
Based on   radial velocities yielded by LAMSOT spectra 
and visual examination of the SDSS images, we find 28 objects, 
 5 bona fide and 23 likely GCs. 
Amongst the five  bona fide GCs, three have been recently
discovered independently  by others, the remaining 25 are our new identifications, 
including two {bona fide} ones.
The new identified objects fall at projected distances ranging from
13 to 265\,kpc from M31.   
Of the two newly discovered bona fide GCs, one is  located near M33, probably  a GC belonging to M33.
The other  bona fide GC  falls on the Giant Stream with a projected distance of  78\,kpc from M31. 
Of the 23 newly identified likely GCs, one has a projected distance
of about 265\,kpc {from M31} and could be an intergalactic cluster. 
}

   \authorrunning{B.-Q. Chen et al. }            
   \titlerunning{M31 GCs from LAMOST}  

   \maketitle

\section{Introduction}

Globular clusters (GCs) are excellent tracers  for the studies of galaxy formation and evolution. 
Stars in a given GC are supposed to
form almost simultaneously from gas of the same chemical composition. Their 
integrated colors and spectra can be  relatively easily interpreted by a Simple Stellar Population
(SSP; \citealt{Renzini1988}). 
GCs are among the intrinsically brightest objects in a galaxy, making them 
more easily observable than stars for a nearby galaxy. 
Most regular large galaxies host a large number of GCs.
They are one of the oldest stellar populations known in a galaxy
and thus they contain important information with regard to the earliest
assemble age history of a galaxy (see \citealt{Brodie2006} for a review). 

The GCs in the Andromeda galaxy M31 are of particular interest. M31 is the most luminous member of the Local Group 
of galaxies, as well as the nearest archetypical spiral galaxy. It serves as one of the best astrophysical laboratories for the studies 
of the physical and astrophysical processes that govern the morphology, kinematics and chemistry, as well as the formation 
and evolution of galaxies. M31 owns an abundant population of GCs, 
much larger than the Milky Way (MW). \citet{Barmby2001} estimate that M31 probably have $\sim$475$\pm$25 
GCs, about 3 times that of the MW (157 GCs; \citealt{Harris1996}). This ratio of GC numbers of 
M31 and the MW has recently increased to more than 4 
as M31 has been more thoroughly searched. In total, 638 objects are classified as
confirmed GCs in the Revised Bologna Catalog of M31 GCs and candidates, Version 5 
(RBC V5; \citealt{Galleti2004}  and most recently updated in August, 2012). More GCs and candidates are 
identified in recent years. \citet{diTull2014} find 7 objects, including 6 confident GCs and 1 
candidate, in the M31 halo by visual examination of the 
Sloan digital Sky Survey (SDSS) images, 
excluding background galaxies based on a combination of the optical, ultraviolet, and infrared colors
 of the objects, 
as well as their photometric redshifts deduced from the SDSS data. \citet{Huxor2014} discover
59 GCs and 2 candidates in the halo of M31 again,  via visual inspection of the MegaCam images of the 
Canada-France-Hawaii Telescope (CFHT) collected by  the Pan-Andromeda Archaeological Survey (PAndAS).

At present, there are  $\sim$ 700 confirmed GCs in M31, and hundreds of candidates that need to be further
checked. While many studies of GCs in M31 are based on imaging  surveys \citep{Hubble1932, Sargent1977, 
Crampton1985, Battistini1987,  Kim2007, Peacock2010, Ma2015}, spectroscopic observations 
can provide vital kinematic information and chemical composition of GCs.
The first spectroscopic observations of M31 GCs was presented by \citet{vandenB1969}. This was followed by a large 
number of studies, e.g. as some of the most recent examples, \citet{Ashman1993, 
Barmby2000, Perrett2002, Galleti2006, Lee2008, Caldwell2009, Velj2014} and references there within. 
A large, systematic spectroscopic survey of GCs in M31 that covers from the disk to the outer halo,
is however still lacking.

Most of the previous studies of the M31 GCs concentrate on the disk and inner halo of M31, 
typically within a projected distance  $R_{\rm p} < 30\,$kpc (e.g. \citealt{Crampton1985, Battistini1987, Brodie1991,
Barmby2000, Perrett2002, Kim2007, Galleti2006, Fan2008, Caldwell2009}). More recent studies 
have shifted their focus to the more distant  GCs in the outer halo of M31, out  to a 
projected distance  of $\sim$150 kpc \citep{Galleti2007, Huxor2008, Richardson2011, Tanvir2012, diTull2013, Mackey2013,
diTull2014, Huxor2014, Velj2014, Sakari2015}. A large fraction of those outer halo GCs of M31 are likely assembled via
accretion of cluster-bearing satellite galaxies, as in the case of the MW \citep{Mackey2013, Yuan2013, Sakari2015}. These 
remote GCs thus serve as excellent signposts to search for the remnants and debris  
of tidally disrupted galaxies. Hitherto,  more than 80 GCs of  projected distances greater than  30\,kpc
 have been identified, mostly by the PAndAS survey (\citealt{Huxor2014} and references there within).

The Large Sky Area 
Multi-Object Fiber Spectroscopic Telescope (LAMOST, also named the 
Guoshoujing Telescope;  \citealt{Cui2012})\footnote{http://www.lamost.org/website/en} is a quasi-meridian reflecting 
Schmidt telescope with an effective light-collecting aperture of about 4\,m and a field of view of 5$\degr$ in diameter.  
It is equipped with 4,000 robotic fibers and thus can record spectra of up to  
4000 celestial objects simultaneously. After a two-year commissioning phase and one-year Pilot Surveys, the  
LAMOST Regular Surveys began in October, 2012. The large number of fibers in a wide field of view 
makes LAMOST an ideal facility to carry out a systematic spectroscopic survey of known GCs and candidates 
in M31 and M33, and, at the same, to search and identify new ones. Since the early phase of LAMOST 
operation, as parts of the LAMOST Spectroscopic Survey of the Galactic Anti-center 
(LSS-GAC; \citealt{Liu2014, Yuan2015}),
we have 
use the LAMOST to carry out a systematic spectroscopic campaign of GCs and candidates in the vicinity fields of 
M31 and M33. In addition to GCs and candidates, other interesting objects targeted in this area include planetary 
nebulae (PNe) and background quasars, as well as H~{\sc ii} regions and supergiants in M31 and M33.  
Some of the early results from this campaign have published, see \citet{Yuan2010} for planetary nebulae and 
\citet{Huo2010, Huo2013} for background quasars. For GCs, our purpose is two-fold -- to build up a large, 
systematic spectroscopic dataset for GCs and candidates in this sky area, and to search for new ones. 
In this paper, we will present our first effort of searching for GCs in the  outer halo of M31, 
using the LSS-GAC data accumulated hitherto. Our GC candidates are selected from the SDSS photometric data.
In this paper, we will also present 
a catalog of all GCs and candidates targeted by LAMOST hitherto near M31 and M33. 
Detailed kinematic and chemical analysis of this large sample of GCs and candidates 
will be presented in separated papers (Chen et al. 2015, in preparation). 

This paper proceeds as follows. In \S{2} we describe our target selection.
A brief description of the observation and data reduction is presented in \S{3}. In \S{4} we introduce our 
strategy used to identify new GCs. We present and discuss the results in \S{5}.
Finally a summary  is given in \S{6}.

\section{Target selection}

As part of the LSS-GAC, objects  targeted in the M31 and M33 region include 
several classes of objects reachable by LAMOST at the distance of M\,31 
(about 770\,kpc), such as GCs, PNe, \HII~ regions
and supergiants and background quasar. For a given plate, 
spare fibers are filled with  foreground Galactic 
stars \citep{Yuan2015}. We hereby introduce the GC target selection of LSS-GAC.
The objects  are primarily selected from  two sources, existing
M31 GCs and candidates,  and new candidates selected from the SDSS photometric catalogs. 
In this paper, we refer to  all the targets
selected from the literature as the `Literature sample' and all those selected from the SDSS 
photometry as the `SDSS sample'.
For all types  of targets  included  in target input catalogs of LSS-GAC, 
GC targets are almost always assigned  the highest priority (prior = 2).

\subsection{Revised Bologna Catalog of M31 GCs and candidates}

To collect all known GCs and candidates in M31, we choose
RBC\footnote{http://www.bo.astro.it/M31/. RBC has been updated to Version 5 since August, 2012.
At the time of our  target selection (2010), it was of Version 4 published  in December, 2009.} 
\citep{Galleti2004} as our starting point. RBC is a main repository 
for information of the M31 GC systems. It is a compilation of
previously published catalogs. The catalog lists all the confirmed GCs (classes 1 and 8 )
 and candidates (classes 2 and 3) at the
time that the original catalog was published, and also all the objects that were 
originally identified as GC candidates and then 
subsequently recognised not to be genuine clusters, such as  \HII~ regions (class 5), 
background galaxies (class 4) or foreground stars (class 6). The identities of these  
contaminants are retained in the RBC in order to avoid their re-discoveries 
as M31 GCs.
Before the start of each observational season (in September usually), 
all published newly discovered GCs in the M\,31 and M\,33 area are added 
to the input source catalogs of LSS-GAC for the preparation of observational plates.

\subsection{New Candidates selected from the SDSS photometry}

\begin{figure}
\centering
   \includegraphics[width=0.58\textwidth]{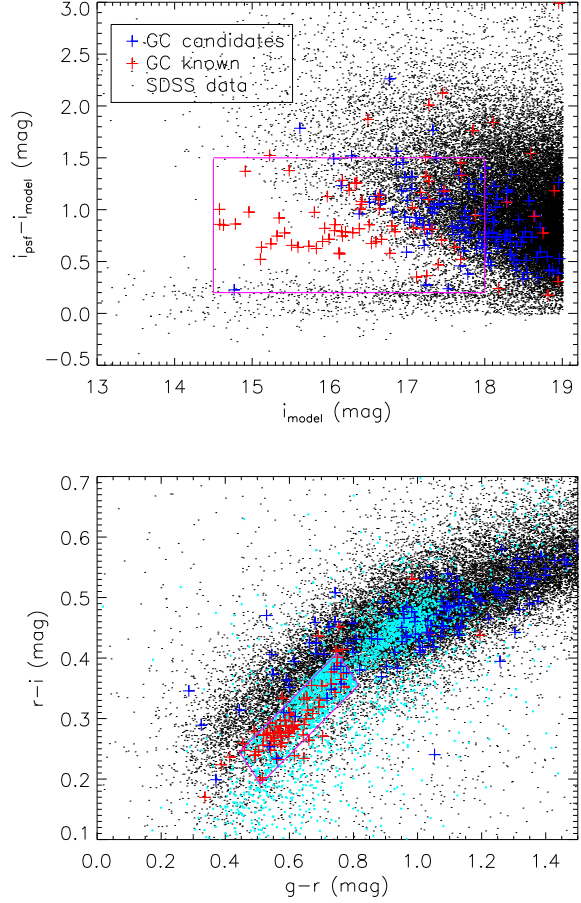}
\caption{{\it Upper panel}: Differences of i-band PSF and model magnitudes plotted against the i-band  model
magnitude for SDSS sources of the M31 stripes; {\it Lower panel}: $g-r$ versus
$r-i$ color-color diagram of the same sources. In both panels, the red and blue pluses
are respectively known GCs and GC candidates from \citet{Peacock2010}. The pink
lines delineate the area of potential  candidates (see text). The cyan dots in the lower
panel denote our  final selected GC candidates.} 
\label{sdssgal}
\end{figure}

The Sloan Digital Sky Survey (SDSS; \citealt{York2000}) has observed a large sky area around M31.
The SDSS photometric catalogs provide us the possibility of selecting interesting sources such as 
GC candidates in the vicinity fields of  M31 and M33. Our target selections were based on the SDSS Data Release 8 (DR8;  
\citealt{Aihara2011}). The sky coverage of SDSS is incomplete around the area of M31. This unfortunately 
limits the extent of region of our target selections. The GC candidates 
are selected via the following steps.
 
\begin{enumerate}
 \item The SDSS criteria for star-galaxy separation perform well, at a reliability level better than 90\% 
up to $r$ = 21.6\,mag. The performance however degrades 
 at fainter magnitudes \citep{Abazajian2003}. The absolute magnitudes $M_V$
of GCs range between $-10.5 < M_V < -3.5$\,mag \citep{Huxor2008}. Assuming a 
distance of 770\,kpc for M31 \citep{Caldwell2011}, the apparent magnitudes of 
GCs in M31 in SDSS $i$-band range between  $13.5 < i <19.0$\,mag. 
We discard all point sources and select GC 
candidates only from objects classified as non-stellar by SDSS. 
As GCs are red objects, the SDSS $i$-band magnitudes are used to apply an 
apparent magnitude cut for GC candidates. Considering the limiting magnitude of LSS-GAC
survey \citep{Yuan2015}, the magnitude cut is set at $14.5 < i <18.0$\,mag. 

 \item The GC sample of \citet{Peacock2010} is 
used to help create the following GC selection criteria. The sample contains 571 confirmed GCs and 373 
GC candidates. The catalog is cross-matched with the SDSS photometric  
catalogs for extended objects (galaxies), with a matching radius of 3$''$.  This yields a test sample of 83 genuine GCs and 127 GC 
candidates with SDSS photometry. 

The differences between the $i$-band point spread function (PSF) magnitudes  ($i_{\rm psf}$) and
 model magnitudes ($i_{\rm model}$) are used to exclude point sources, such as the foreground stars, as well as 
 those very extended sources, such as the background galaxies and dwarf galaxies.
The upper panel of Fig.~\ref{sdssgal} plots  values of $i_{\rm psf}-i_{\rm model}$ 
versus $i_{\rm model}$ for all SDSS non-stellar objects around the area of M31. 
The confirmed GCs and candidates  from \citet{Peacock2010}  are overplotted.
Most of the confirmed GCs fall  in the range $0.2 < i_{\rm psf}-i_{\rm model} < 1.5$\,mag. This
is adopted as our second selection criterion of GC candidates. Only three confirmed GCs have
$i_{\rm psf}-i_{\rm model}$ values smaller than but close to  0.5\,mag. We set the lower limit
of $i_{\rm psf}-i_{\rm model}$ to 0.2\,mag, in order  to include as many GC candidates as possible.
  
\item GCs have relatively narrow color ranges \citep{Peacock2010}. Thus a  color cut can also help to separate  
GCs from galaxies and stars. Only colors $g-r$ and $r-i$ are used, 
given the better signal-to-noise ratios (SNRs) of  SDSS photometry   $g$-, $r$- 
and $i$-bands than in $u$- and $z$-bands. 
A color-color diagram of all SDSS sources, the selected candidates, 
together with the known GCs and candidates from  \citet{Peacock2010}  is presented in the lower panel of 
Fig.~\ref{sdssgal}. Most of the known GCs are located in a small area in the color-color diagram, except for a few outliers. 
Based on this  plot, we have therefore defined the following  color cuts for selecting GC candidates,
 \begin{eqnarray*}
0.75<(g-r)+1.25(r-i)<1.26{\rm \,mag} \\
-0.08<(r-i)-0.53(g-r)<0.01{\rm \,mag}        
 \end{eqnarray*}
In the current work, we only adopt the above color cuts to objects fainter than an $i$ magnitude of 16.0\,mag, 
considering that  few background
galaxies can reach an apparent $i$ magnitude of  brighter than  16.0\,mag.
\end{enumerate}

We obtain a sample of 3,585 candidates from the criteria above in the vicinity fields of M31 and M33. They are overplotted 
as cyan points in the bottom panel of Fig.~\ref{sdssgal}. The sample is cross-matching with  the existing catalogs
such as RBC. Duplicates  are discarded. Finally, 3,280 candidates remain.
Together with GCs and candidates known in the literature, a total of  more than 6,000 GC targets are included in the 
input  catalogs of LSS-GAC.
We are aware that  the sample is likely to contain many contaminants, such as background
galaxies and quasars. We have adopted a relatively
loose set of selection criteria, taking advantage
the top spectral collection rate of LAMOST.

\section{Observations and Data reduction}

\begin{figure*}
\centering
   \includegraphics[width=0.8\textwidth]{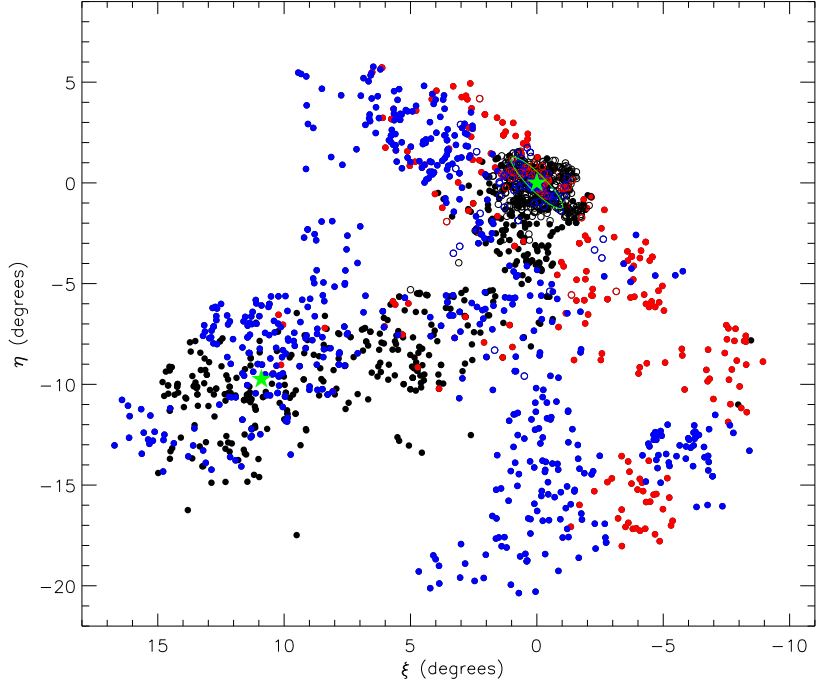}
\caption{Spatial distribution of GC candidates selected from the SDSS  photometry (`SDSS sample') and targeted by LAMOST
in 2011(Black filled circles), 2012 (red filled circles) and 2013 (blue filled circles. 
Targets from the literature (`Literature sample') targeted in those three years are represented by open circles 
of the corresponding colors. For objects targeted  multiple times, the color coding refers to  
the observation with the highest \bsnr. 
The green stars mark the central positions of M31 
and M33, respectively. The green ellipse represents the optical disk of M31 of radius 
$R_{25}\,=\,95'.3$.} 
\label{seldis}
\end{figure*}

\begin{table}
 \centering
  \caption{LAMOST observations of GC targets  in the vicinity fields of M\,31 and M\,33.}
  \begin{tabular}{lcccc}
  \hline
  \hline
Observational   & Observing   &  No. of   & \bsnr~  & \bsnr~  \\
  season  & nights       &  Spectra  & $>$5      & $>$10   \\
 \hline
2011.9--2012.6  &  41  & 3292 & 1166  & 589 \\
2012.9--2013.6  &  14  & 601  &  289    & 184 \\
2013.9--2014.6  &  26  & 865  &  503    & 322 \\
 \hline
\end{tabular}\\
  \label{ta1}
\end{table}

Following a two-year commissioning phase, the LAMOST Pilot 
Surveys were initiated in October 2011 and completed in June 2012. 
The Regular Surveys, expected to last five years,  were initiated in October 2012. 
In the current work we present results based on data collected by
LAMOST in the 2011, 2012 and 2013 observational seasons, 
i.e. from the Pilot Survey and the first two years of the Regular Surveys.

Being a quasi-meridian reflecting Schmidt telescope, LAMOST only observes a given 
field of plate between 
2\,h before and after the transit. The M31 and M33 area  $(0<{\rm RA} <30\degr, 
25<{\rm Dec.}<50\degr)$ are  targeted by LAMOST from September to 
January of the next year in a given observational season, 
which starts in September and ends in June of the following year. 
The plates are observed in nights  of dark or grey lunar conditions. 
Typically 2--3 exposures are obtained  for each 
plate, with an integration time per exposure varying, depending on  the weather conditions, 
between  600--1200\,s, 1200--1800\,s and 1800--2400\,s for bright (B),  median (M) and 
faint (F) plates, respectively. Some observing nights reserved to monitor the telescope performance 
are also used to observe M31 and M33 plates. For most plates, the seeing varies between 3 --
4\,arcsec, with a typical value of about 3.5\,arcsec \citep{Yuan2015}. 

LAMOST has a field of view (FoV) of $5\degr$ in diameter.
There are 16 low-resolution spectrographs, each accommodating 250 auto-positioning fibers. 
The  parking positions of  fibers evenly distributed in the focal plane, 
except for a few regions reserved for guiding cameras and the central Shack-Hartmann sensor. 
Each fiber has a diameter of 
3.3\,arcsec projected on the sky. LAMOST is implemented with slit
masks of width 2/3 the fiber diameter, i.e. 2.2\,arcsec, yielding a spectral resolving 
power of about $R \sim 1800$. The spectra cover a wavelength range of 
3700--9000\,\AA. The light entering each spectrograph is dispersed and recorded in two arms,
covering 3700 -- 5900\,\AA~  and 5700 -- 9000\AA~ in the  blue and in the red, respectively.
In each arm of a given spectrograph, a $4096 \times 4096$ CCD, with a squared pixel size 
of 12\,$\rm \mu$m, is used to record the light signal \citep{Cui2012}. One CCD pixel corresponds 
to about 0.56\,\AA~ and 0.82\,\AA~ in the blue and in the red, respectively.

We give a summary of the observations of GC targets observed by LAMOST in the M31 and M33 area  in Table~1.
By June 2014, a total of 131 plates with GC targets have been collected, yielding 4,758 spectra of 1,991 unique GC. 
Most of the  spectra (3292) were observed in the 2011 observational 
season, while  601 and 865 spectra were collected in the 2012 and 2013 seasons, respectively.  About 
42.2, 25.1, 12.0, 7.9, 4.9, 4.0 and  2.3 per cent objects were targeted  by 1 to 7 times, respectively. 
There are 1958, 1095 and 465 spectra having \bsnr~  greater than 5, 10 and 20  per pixel, 
respectively. As GCs are red objects, the SNRs are better in the red than in the blue. 
This is reflected in the fact that there are 
3101, 2398 and 1488 spectra that have  \rsnr~ greater than 5, 10 and 20 per pixel, respectively. 

Fig.\ref{seldis} plots the spatial distribution in the $\xi$ - $\eta$ plane of all GC targets observed by LAMOST by 
June 2014 in the vicinity fields of M31 and M33 (located at $\xi = 11\degr.3,~ \eta = -10\degr.1$ on this map).
Here $\xi$ and $\eta$ are respectively the offsets in  Right Ascension and Declination relative to the optical center of M31 
(RA=00$^{\rm h}$42$^{\rm m}$44.30$^{\rm s}$; Dec=$+$41\degr16$^\prime$09.0$^\prime\prime$; 
 from \citealt{Huchra1991}; \citealt{Perrett2002}). 
The green ellipse in Fig.\,2 represents the 
optical disk of M31, with an optical radius $R_{25}$ = 95$^\prime$.3 \citep{deVau1991}, an inclination angle $i$ = 77$\degr$ 
and a position angle ${\rm P.A.}$ = 38\degr \citep{Kent1989}.  
Candidates of the SDSS Sample are mostly located in the outer 
halos of M31 and M33 , while those of the Literature Sample fall mostly  near the disk
and inner halo of M31.

\begin{figure}
\centering
   \includegraphics[width=0.48\textwidth]{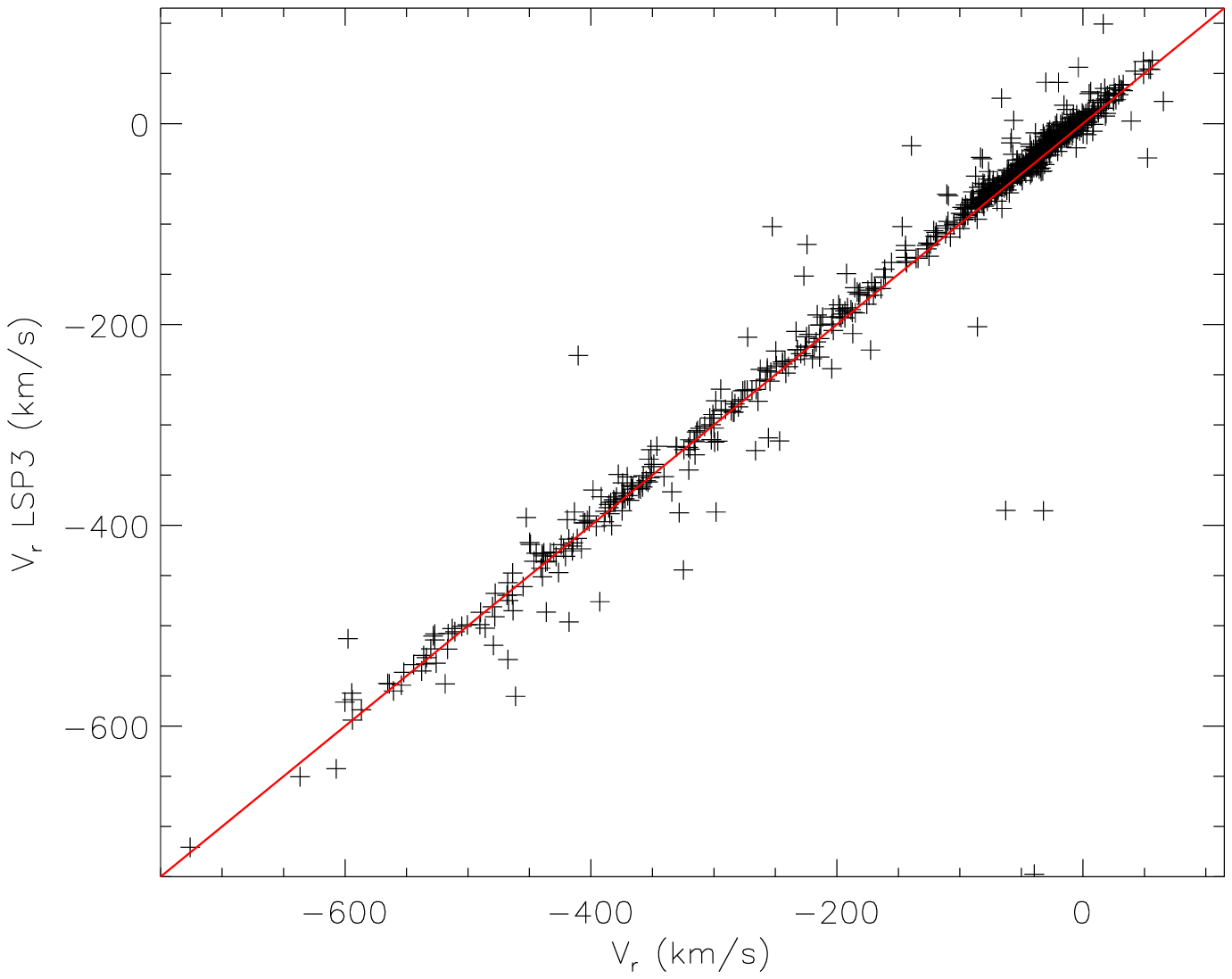}  
   \includegraphics[width=0.48\textwidth]{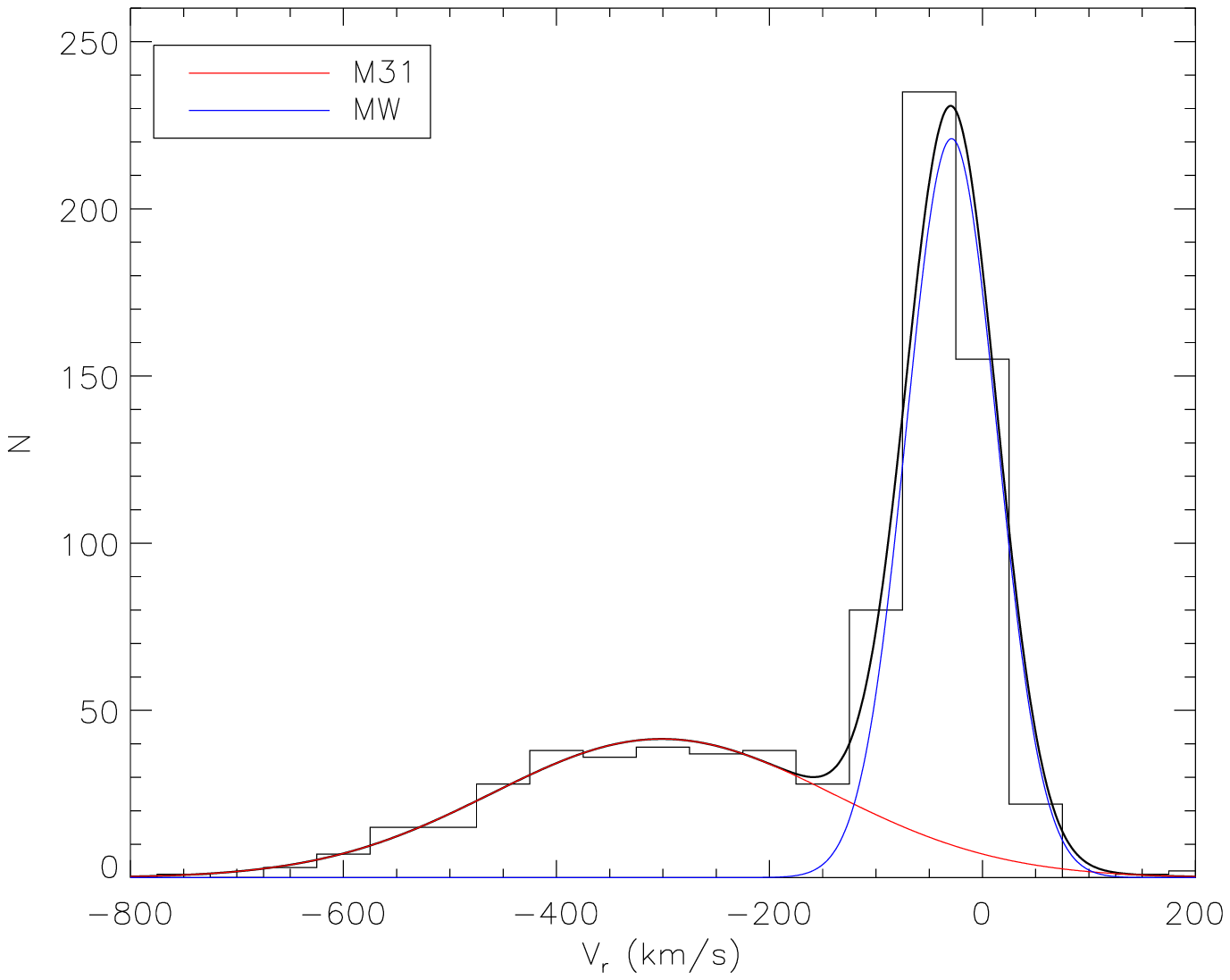}
\caption{{\it Left panel}: Radial velocities derived in the current work for all non-galaxy candidates that have
\bsnr~$>$5 per pixel, compared to the values yielded by the LSP3 \citep{Xiang2015b}. 
A red line denoting complete equality is overplotted to guide the eyes. 
{\it Right panel}:  Histogram distribution of
newly derived radial velocities for the same set of objects.
The means and standard deviations of Gaussians overplotted are taken from 
\citet{Galleti2006}. } 
\label{rvdist}
\end{figure}

All the spectra were first processed with the LAMOST 2-D pipeline (Version 2.6, 
\citealt{Luo2015}), including steps of bias subtraction, cosmic-ray removal, 
1D spectral extraction, flat-fielding, wavelength calibration, and sky subtraction. 
he blue- and red-arm spectra are
processed separately in the 2D pipeline and then joined together after flux calibration, 
which is carried out using a pipeline specifically designed for the LSS-GAC survey 
\citep{Xiang2015a}. No scaling or shifting is performed in cases where the blue- 
and red-arm spectra are not at the same flux level in the overlapping region, as it is unclear
whether the misalignment is caused by poor flat-fielding or sky subtraction, or both. 
Note that in the current implementation of flux-calibration of \citet{Xiang2015a}, 
the telluric absorptions, including the prominent Fraunhofer A band at
7590\,\AA ~and B band at 6867\,\AA~ have not been removed.
Finally, for objects that have been observed for multiple times in different plates, 
spectra of lower \bsnr~ are scaled by low order polynomials 
to match continuum level of the spectrum of highest \bsnr~
and then combined all together, with each  spectrum weighted by the 
inverse square of errors.

For stars observed in the  LSS-GAC, stellar parameters deduced from the spectra 
with the {LAMOST Stellar Parameter Pipeline (LASP}, \citealt{Luo2015}), 
including radial velocities and basic stellar atmospheric parameters
($T_{\rm eff}$, log$g$ and [Fe/H]), are available from the LAMOST DR1 \citep{Luo2015}. A separate
pipeline, the LAMOST Stellar Parameter Pipeline at Peking University (LSP3; \citealt{Xiang2015b}), 
has been developed at Peking University to determine the radial velocities and values of $T_{\rm eff}$, 
log\,$g$ and [Fe/H]. The LSP3 determines the stellar atmospheric parameters by spectral template 
matching with the MILES empirical spectral library \citep{Sanchez2006, Falcon2011}, 
instead of the ELODIE library \citep{Prugniel2001, Prugniel2007} used by 
the LAMOST default pipeline. The LSP3 is applied to 
all spectra having  a \bsnr~ $>$ 2.76 per pixel, including GC targets of interest here.
The LSP3 however treats the  GC spectra as of stars.

For our GC targets, we have thus derived  their radial velocities, as well as the ages and metallicities by  comparing 
the observed spectra with those of SSP models  (Chen et al. 2015, in prep.), using the public code 
ULySS\footnote{http://ulyss.univ-lyon1.fr/} (University of Lyon Spectroscopic analysis Software; \citealt{Koleva2009}). 
ULySS is an open-source software package used to study the stellar populations of galaxies star clusters, 
as well as to derive the  atmospheric parameters of stars. 
It performs spectral fitting with a linear combination of non-linear components, 
convolved with a line-of-sight velocity distribution and multiplied by a polynomial continuum.
For likely GC candidates in our sample, we use the routines to fit the spectra  and determine the target properties. 
SSP models computed with the MILES library \citep{Vazdekis2010} are adopted. 
The MILES spectra cover the wavelength range 3540--7400\,\AA~ at a resolution 
of 2.5\,\AA FWHM. Only spectra having a \bsnr~ $>$ 5 or \rsnr~ $>$ 10 per pixel  are analyzed. 
Spectra that have a large number of bad pixels ($>$ 1/3 of the total) are also excluded. This leaves
us with {908} unique objects in our sample.  
The radial velocities are derived if they fall in the range $-2000<V_r<2000$\,\kms. Background
galaxies, typically having a radial velocity $V_r > 5000$\,\kms, are first excluded by visually examination before fitting  
the observed spectra  with models. 

By internal tests using multiple observations of duplicate objects and 
external tests using common targets with previous studies in the literature, 
we conclude that for GC targets with  \bsnr~ $>$ 5, the derived radial velocities have achieved 
an accuracy of better than  12\,\kms~ (Chen et al. 2015, in prep.). 
In the left panel of Fig.~\ref{rvdist}, we compare our newly derived radial velocities with those delivered by the 
LSP3 \citep{Xiang2015b} for all the non-galaxy objects that have \bsnr~ $> 5$. 
They  are in very good agreement, 
with a marginal system offset  and scatter of $V_{r}-V_{r,{\rm LSP3}}= -5 \pm 6\,$\kms. 
A histogram  distribution of the newly derived radial velocities for the same set of objects  is 
presented in the right panel of Fig.~\ref{rvdist}.  A double  two Gaussian distribution, with one 
peak at around $-30\,$\kms representing the MW foreground stars, and another around $-300$\,\kms~
for M31 objects, is clearly visible. Also overplotted are two Gaussians,
with the parameters taken from \citet{Galleti2006}. The first Gaussian has a  mean $\mu = -301.0\,$\kms~
 and a standard deviation $\sigma$ = 160.0\,\kms~ for the M31 GCs, 
and the second one has a  mean $\mu = -29.0$\,\kms~ and a standard deviation $\sigma$ = 42.6\,\kms~ for
the MW foreground stars. The sum of the two Gaussian functions fit nicely the velocity distribution. 
The Figure shows  that there
is a large number  of  foreground MW stars contaminating the  sample.

\section{The GC identifying}

\begin{figure*}
\centering
   \includegraphics[width=0.98\textwidth]{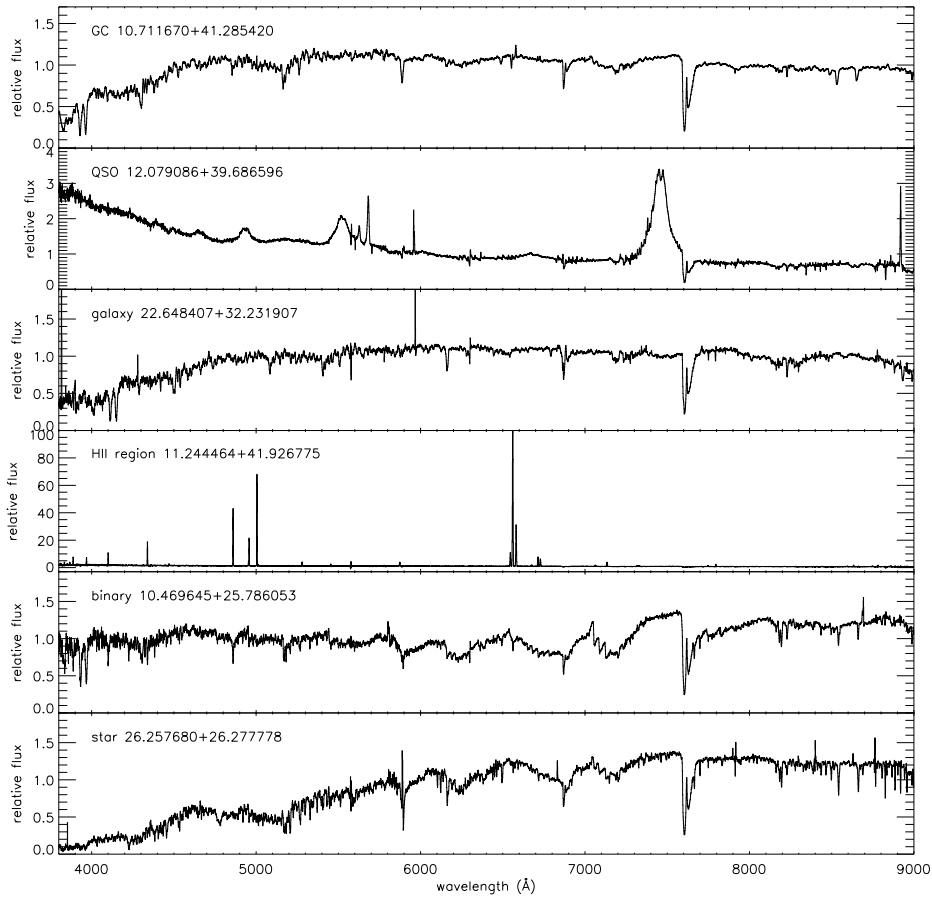}  
\caption{Example LAMOST spectra of  targets classified as non-GC objects, {as well as that of  a known GC}. 
From top to bottom, the Figure plots, respectively, {the LAMOST}
spectra of {a known GC, a} quasar, a galaxy, an \HII~ region, 
a white-dwarf-main-sequence binary and an M-type main-sequence star. 
For each spectrum, the  object type and coordinates (RA and Dec. in degree) are labelled.} 
\label{othersp}
\end{figure*}

\begin{figure}
\centering
   \includegraphics[width=0.68\textwidth]{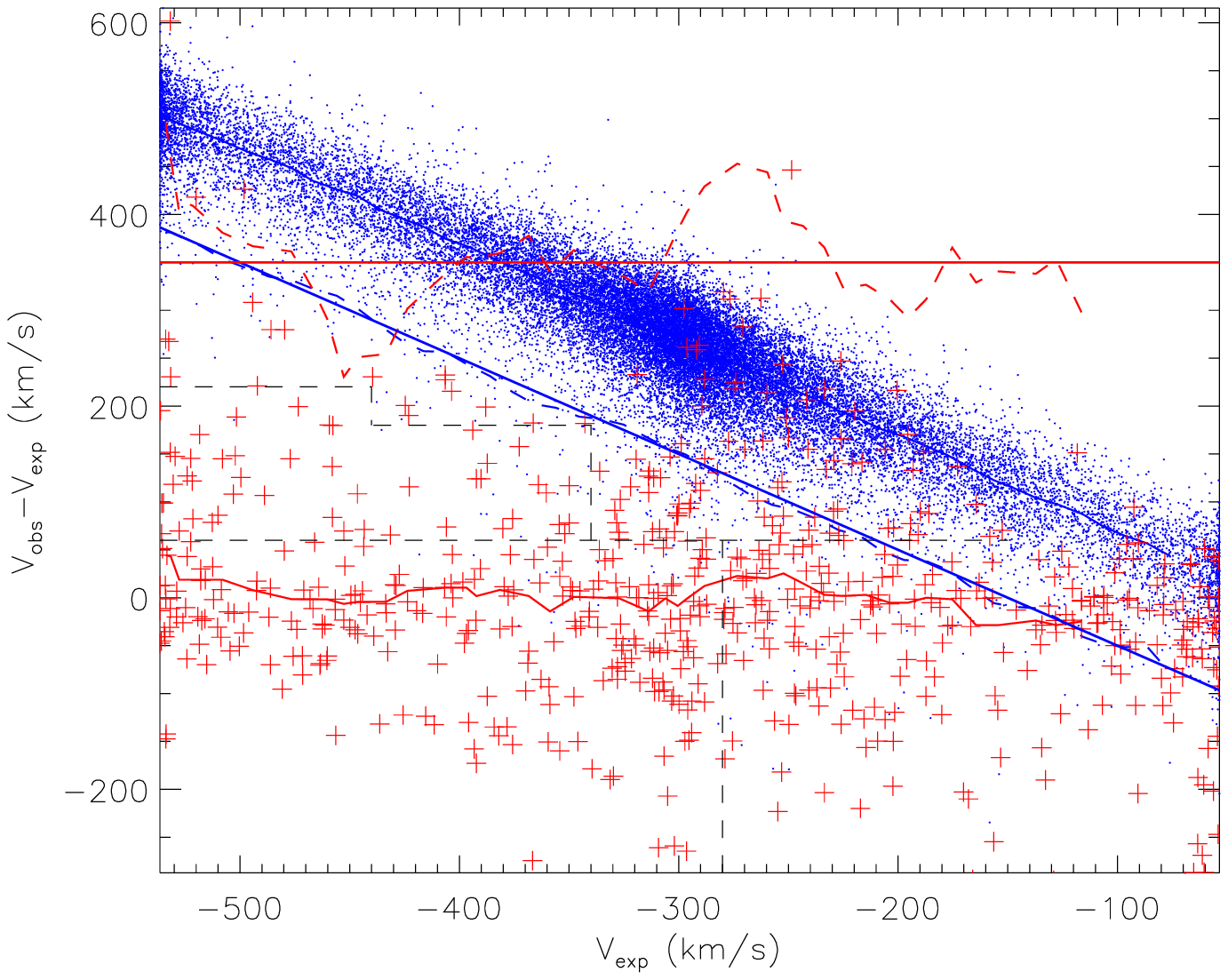}  
\caption{Comparison of the observed and expected radial velocities of known GCs from
the literature (red pluses) and of foreground MW stars simulated with the Besan\c{c}on model  (blue dots). 
Here we plot the differences between the observed and expected radial velocities 
against the expected radial velocities. The blue and red solid lines give the binned median values
for known M31 GCs and the simulated foreground stars, respectively, with the 
3$\sigma$ scatters delineated by the 
blue and red dashed lines, respectively.
The blue solid straight line marks a line of constant  $V_{\rm r} = - 150$\,\kms~ , while  the 
red solid straight line traces a line of  $V_{\rm r}-V_{\rm exp}=350$\,\kms. } 
\label{besm}
\end{figure}

Some of our targets that are clearly not GCs objects can 
be readily singled out based on the  spectra obtained. 
There are 15 objects in our sample whose spectra display strong TiO bands typical of M-type stars. One target shows
a composite spectrum, a blue component of a white dwarf  and 
a red component of an M dwarf, i.e. this  is a white-dwarf-main-sequence binary.
These targets will be classified as stars in our catalog.  There are 8  objects displaying strong and broad emission lines 
characteristic  of quasars, and are  classified as quasars.  Example 
spectra of those classified as non-GC objects are presented in Fig.~\ref{othersp}.

\begin{figure}
\centering
   \includegraphics[width=0.68\textwidth]{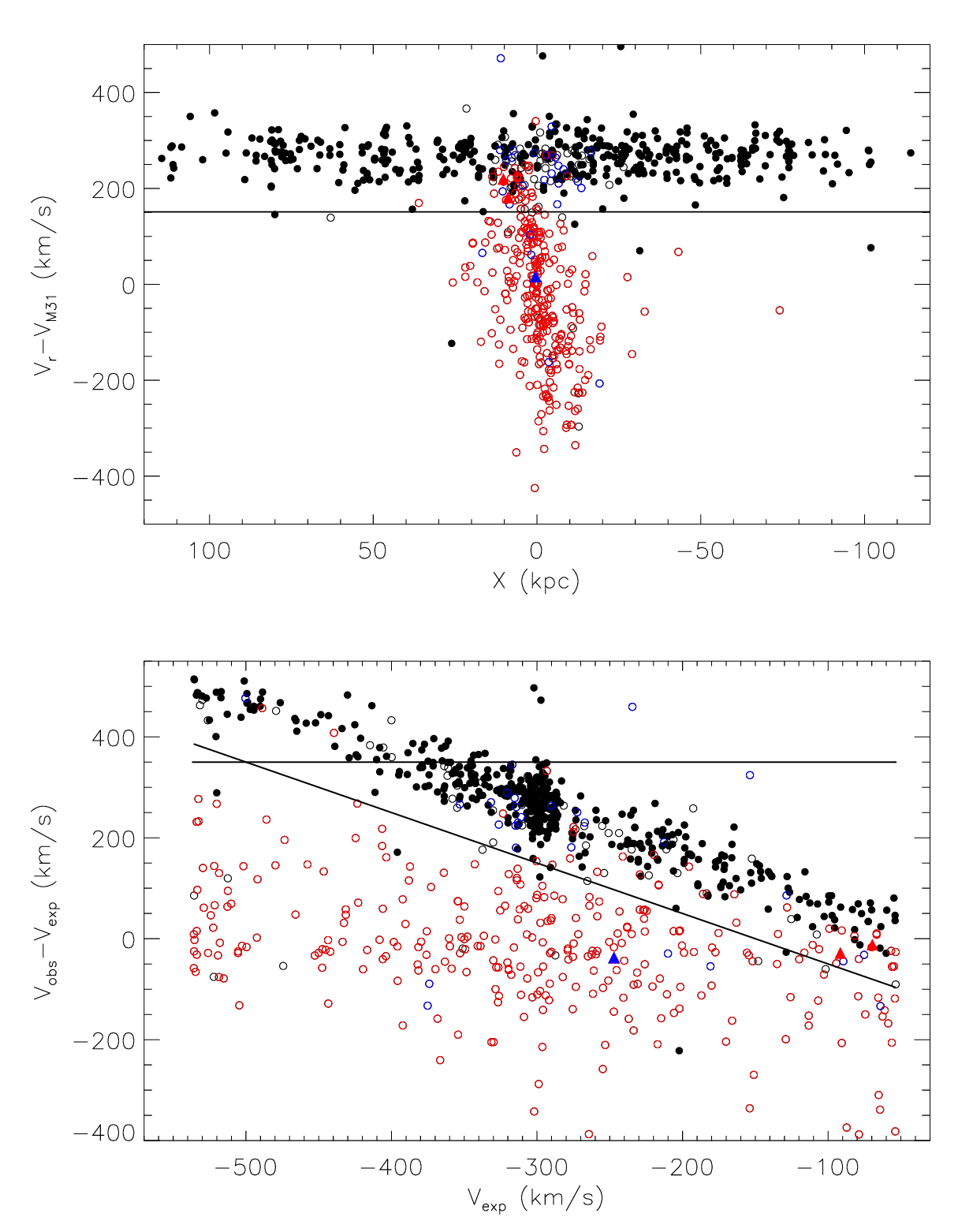}  
\caption{{\it Upper panel}: Velocity offsets relative to the systematic value of M\,31, 
plotted against major-axis distance $X$, for GCs
and GC candidates observed by LAMOST. {\it Lower panel}: Same as Fig.~\ref{besm} but for 
GCs and GC candidates observed by LAMOST. In both panels the filled circles, open circles and filled triangles 
represent targets from the SDSS Sample, and those from RBC and from \citet{Johnson2012}, respectively. 
For the RBC targets, 
known GCs (classes 1 and  8) are marked in red, GC candidates (class 2) in blue and objects of 
other  classes in  black. For the
\citet{Johnson2012} targets, confirmed clusters (flagged by a `c') are marked in red and possible clusters
(flagged by a `p') in blue.} 
\label{velc}
\end{figure}

In addition to the aforementioned contaminants of special spectral characterises, 
 there are two other main groups of contaminants.
One  is the background galaxies and the other the foreground Galactic stars. With the
information of radial velocities, we can easily identify the background galaxies since they generally 
 have very large positive radial velocities, 
in contrast to negative velocities of a mean around $-301$\,km\,s$^{-1}$ and a dispersion about 160\,km\,s$^{-1}$ 
(\citealt{Galleti2006}; see also the 
right panel of Fig.~\ref{rvdist})  for GCs of M31. Amongst our SDSS Sample targets, 
we find 218 background galaxies.

Identifying foreground MW stars  is more difficult.
For this  purpose, the information of radial velocities is first  used. 
For a disk galaxy such as M31,  the radial velocity $V_r$
can be approximated by $V_r = V_0 + V(R)\,{\rm sin}\xi \, {\rm cos}\theta$ \citep{Rubin1970}, 
where $V_0$ is the systemic radial velocity, $\xi$ is the angle between the 
line of sight and the norm of the galaxy disk plane, $V(R)$ is the rotational velocity  in
the plane at a radius $R$. cos$\theta$\,=\, $X/R$, where $X$ is the position 
along the major axis. If we assume the rotation curve has a constant velocity, $V(R)\,=\,{\rm const.}$, 
then there is  a linear relation  between $V_{\rm r}$ and object position. \citet{Drout2009} present a relation to calculate the 
expected radial velocity $V_{\rm exp}$, based on  a least-squares linear regression to the 
 radial velocity measurements  of  \HII~ regions in M31,
\begin{equation}
V_{\rm exp} = −295 + 241.5(X/R)\,{\rm km\,s^{-1}}.
\end{equation}
We collect known GCs in M31 that have radial velocities measured  in the literature: 
RBC V5 \citep[and references there within]{Galleti2006}, 
\citet{Caldwell2009} and \citet{Velj2014}. Radial 
velocities of 663 known M31 GCs are collected. These GCs cover a wide spatial distribution, with
positions along the major axis spanning $-85<X<102\,$kpc and along the minor axis ranging $-118<Y<122$\,kpc. 
The linear relation  
between $X/R$ and $V_{\rm exp}$ given by  Eq.~(1) works well for these known GCs. 

For the foreground Galactic stars, the Besan\c{c}on model \citep{Robin2003} is used
to simulate the velocity distribution of Galactic stars in a 5$\times$5\,sq.deg. region around M31. 
We plot the differences between the 
observed radial velocities of all the known M31 GCs and the expected velocities given by 
Eq.~(1) in Fig.~\ref{besm}. Also overplotted are the differences between the modelled radial 
velocities of  foreground Galactic stars and those given by Eq.\,(1).
All the M31 GCs fall along the line of zero differences
($V_r - V_{\rm exp} \sim 0$\,\kms), while all the simulated foreground Galactic
stars locate along a line of radial velocity $V_r \sim 0$\,\kms. 
Lines delineating 3$\sigma$ scatters of velocity differences for both M31 GCs
and simulated Galactic stars  are also plotted in Fig.~\ref{besm}. Essentially all the Galactic stars
fall above the  straight line of $V_r=-150\,$\kms~, except for a few ($< 2$ per cent) that
have a radial velocity less than $-150$\,\kms. Candidates in our sample that have a radial velocity 
 $V_r < -150$\,\kms  thus have a high probability of being genuine GCs in M31.
The same criterion is used by  
\citet{Galleti2006}  to classify bona fide GCs. Also overplotted in Fig.\,5 is 
the criterion adopted by  \citet{Drout2009} to select M31 stars from the Galactic foreground stars.
The two criteria are almost identical.
Only five (1 per cent)  M\,31 GCs fall above the line of  $V_r-V_{\rm exp} > 350\,$\kms~ 
in Fig.~\ref{besm}, approximately the 3$\sigma$ upper limit  for the known M31 GCs. 
We classify all candidates in our sample that have a radial velocity 
$V_r-V_{\rm exp} > 350\,$\kms  as foreground Galactic stars. 
For sample targets of   $V_r > −150.0$\,\kms~  but  $V_r-V_{\rm exp} < 350\,$\kms,
the probabilities of the targets being  foreground Galactic stars cannot be neglected. These objects are 
therefore classified as possible GC candidates in the current work.
 
The above criteria for classifying  candidates are then applied to all targets in our sample, 
excluding those that have been classified as background quasars, galaxies or objects of 
special spectral characteristics, as described above. . 
We calculate the differences $V_r-V_{\rm exp}$ for the remaining objects
in our sample  and plot the differences against $V_{\rm exp}$ in Fig.\ref{velc}.
The boundaries separating  highly confident GC candidates, 
possible GC candidates and stars are also overplotted.
The  classification leads us  a list of 5 new 
M31 GC candidates and 352 possible candidates from the SDSS Sample, 
in addition to those from the Literature Sample.  

A method similar to that used by  \citet{diTull2013} is then applied to search for genuine GCs amongst  the 
5 candidates and 352 possible candidates from the  SDSS Sample  of targets.
We visually examine the SDSS images to search for morphological evidence of clusters, using the 
cutout images retrieved from the SDSS website\footnote{http://skyserver.sdss.org/dr12/en/tools/chart/listinfo.aspx}. 
Some objects are found to be clearly not clusters and are most likely 
binaries, stars or background galaxy pairs. 
For some objects, the SDSS $r$- band images were retrieved
for a more detailed inspection. We adopt the classification scheme of \citet{diTull2013}. The objects
are classified as GCs and candidates by the same criteria as category 1 for GCs and categories 2 and 3
for candidate clusters, respectively. A large number of the candidates 
cannot be reliably classified given the relatively poor resolution of SDSS images. They are denoted as unknown objects in
our catalog. Finally, we are left with 5 GCs, 23 candidate clusters and 277 unknown objects. 

Using the criteria described above, for the original 553 GC targets selected  from the SDSS photometry, we are left with
\begin{enumerate}
\item 5 M31 GCs of $V_r <−150.0$\,\kms or $V_r > −150.0$\,\kms~ 
but $V_r-V_{\rm exp} < 350\,$\kms, with supporting  
morphological evidence;
\item 23 M31 GC candidates of  $V_r > −150.0$\,\kms~ but $V_r-V_{\rm exp} < 350\,$\kms, with supporting  
morphological evidence;
\item 218 background galaxies of large  radial velocities;
\item 22 stars of $V_r-V_{\rm exp} > 350\,$\kms, with supporting 
 morphological evidence  or spectra information of being stars;
\item 8 quasars with strong broad emission lines;
\item 277 unknown objects of $V_r > −150.0$\,\kms~ but  $V_r-V_{\rm exp} < 350\,$\kms, with
no clear morphological evidence and spectra information for a proper classification.
\end{enumerate}

\section{Result and discussion}

\subsection{Newly discovered GCs in the halo of M31}

\begin{table*}
 \centering
  \caption{Positions and properties of 5 confirmed GCs from the SDSS Sample of targets.}
  \begin{tabular}{lcrccccccccc}
  \hline
  \hline
Name & RA & Dec. & $X$  & $Y$ & $R_{\rm p}$ & $g-r$ & $r-i$ & $i_{model}$ & $M_V$ & $i_{psf}-i_{model}$ & $V_r$  \\
          & (deg) & (deg) & (kpc)  & (kpc) & (kpc) & (mag) & (mag) &  (mag) & (mag)  & (mag) & (\kms)  \\
 \hline
   LAMOST-1 &   12.23263 &   35.56682 &   -49.75 &   -60.34 &    78.21 &     0.77 &     0.39 &    17.85 &    -6.19 &     1.28 &    -55 \\
   LAMOST-2 &   24.07521 &   30.27437 &   -11.51 &  -204.43 &   204.75 &     0.71 &     0.33 &    17.80 &    -6.18 &     0.49 &   -175 \\
   LAMOST-3$^a$ &   11.18990 &   43.44303 &    26.06 &    14.11 &    29.63 &     0.58 &     0.28 &    16.97 &    -7.05 &     1.09 &   -424 \\
   LAMOST-4$^b$ &    9.03580 &   39.29165 &   -31.37 &    -2.75 &    31.49 &     0.56 &     0.26 &    17.12 &    -6.89 &     1.18 &   -230 \\
   LAMOST-5$^c$ &   14.73496 &   42.46061 &    38.06 &   -21.18 &    43.56 &     0.55 &     0.27 &    15.54 &    -8.44 &     1.04 &   -144 \\
 \hline
\end{tabular}
  \label{ta2}
\begin{flushleft}
$^a$ Identified previously by \citet{Huxor2014} as `PAndAS-36' in their Table 1.\\
$^b$ Identified previously by \citet{diTull2014} as `D' in their Table 1.\\
$^c$ Identified previously by \citet{Huxor2014} as `PAndAS-46' in their Table 1. \\
\end{flushleft}
\end{table*}

\begin{figure}
\centering
   \includegraphics[width=0.68\textwidth]{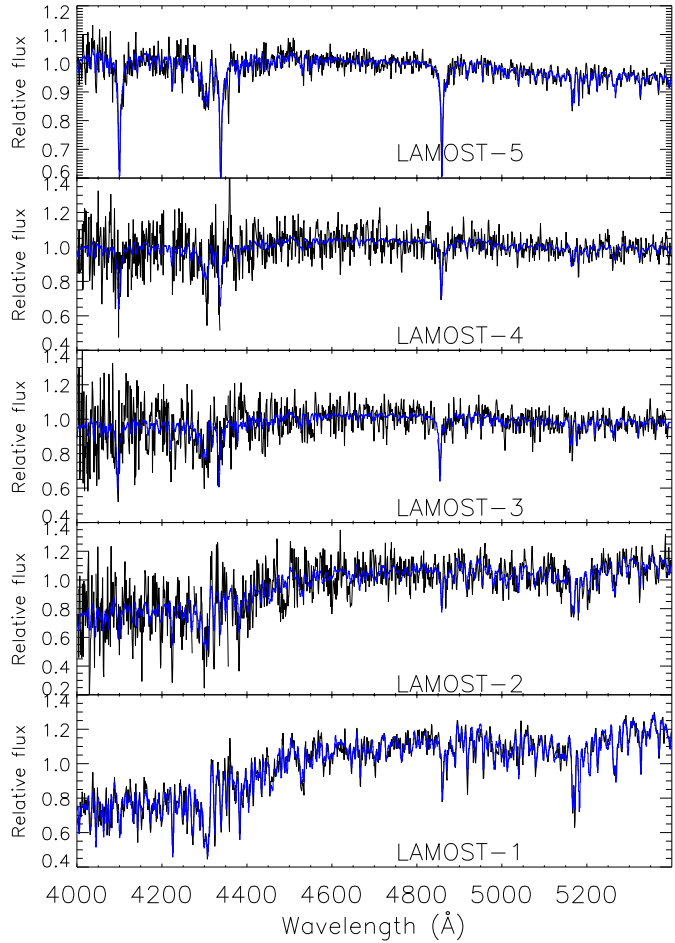}  
\caption{LAMOST spectra of the 5 objects listed in Table~2. The observed and best-model spectra are 
plotted in black and blue, respectively. The plots are labelled with the LAMOST names of the objects.} 
\label{gcsp}
\end{figure}

\begin{figure*}

\begin{overpic}[width=0.315\textwidth]{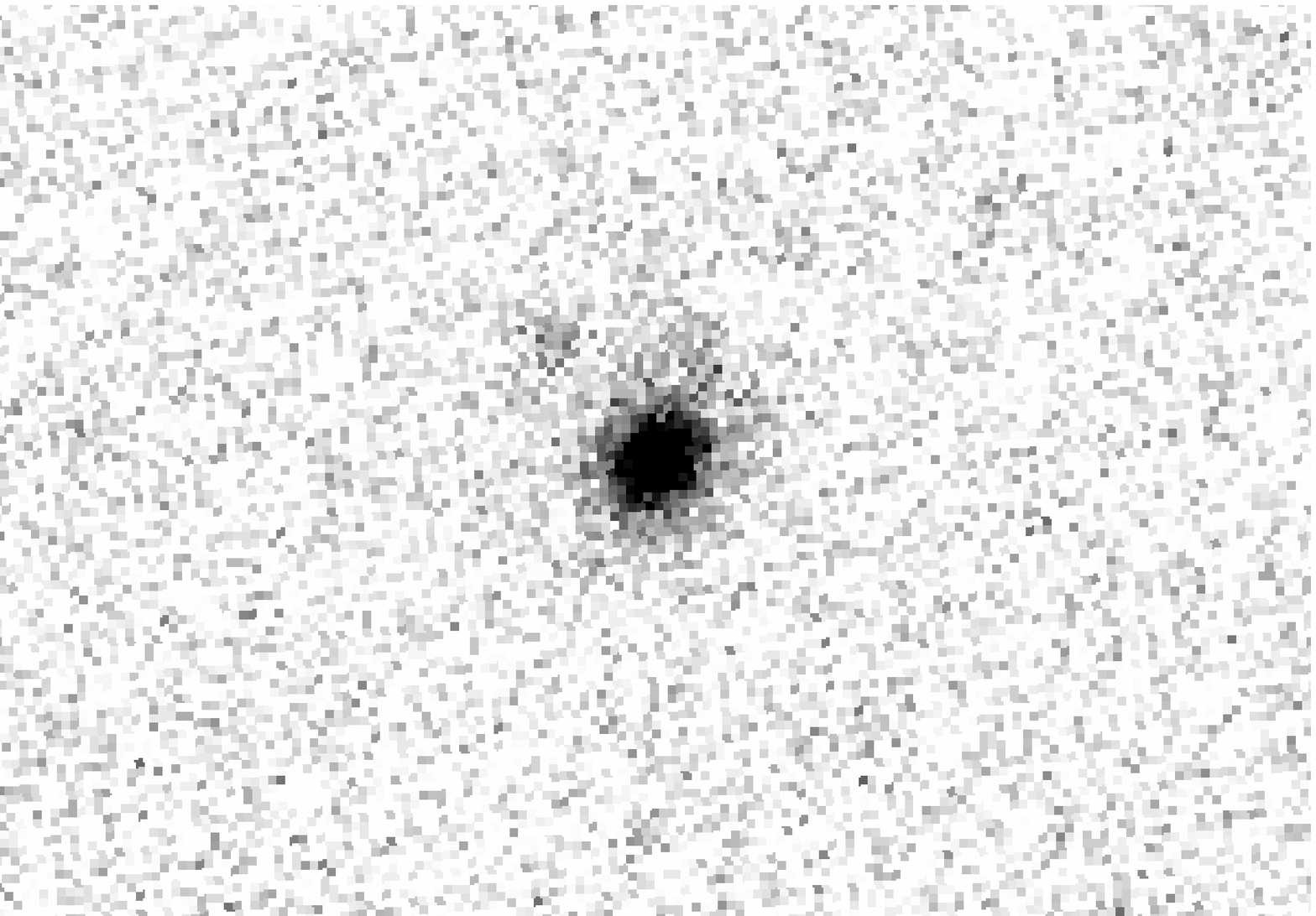}  
\put(3,3){ LAMOST-1}
\end{overpic}
\begin{overpic}[width=0.315\textwidth]{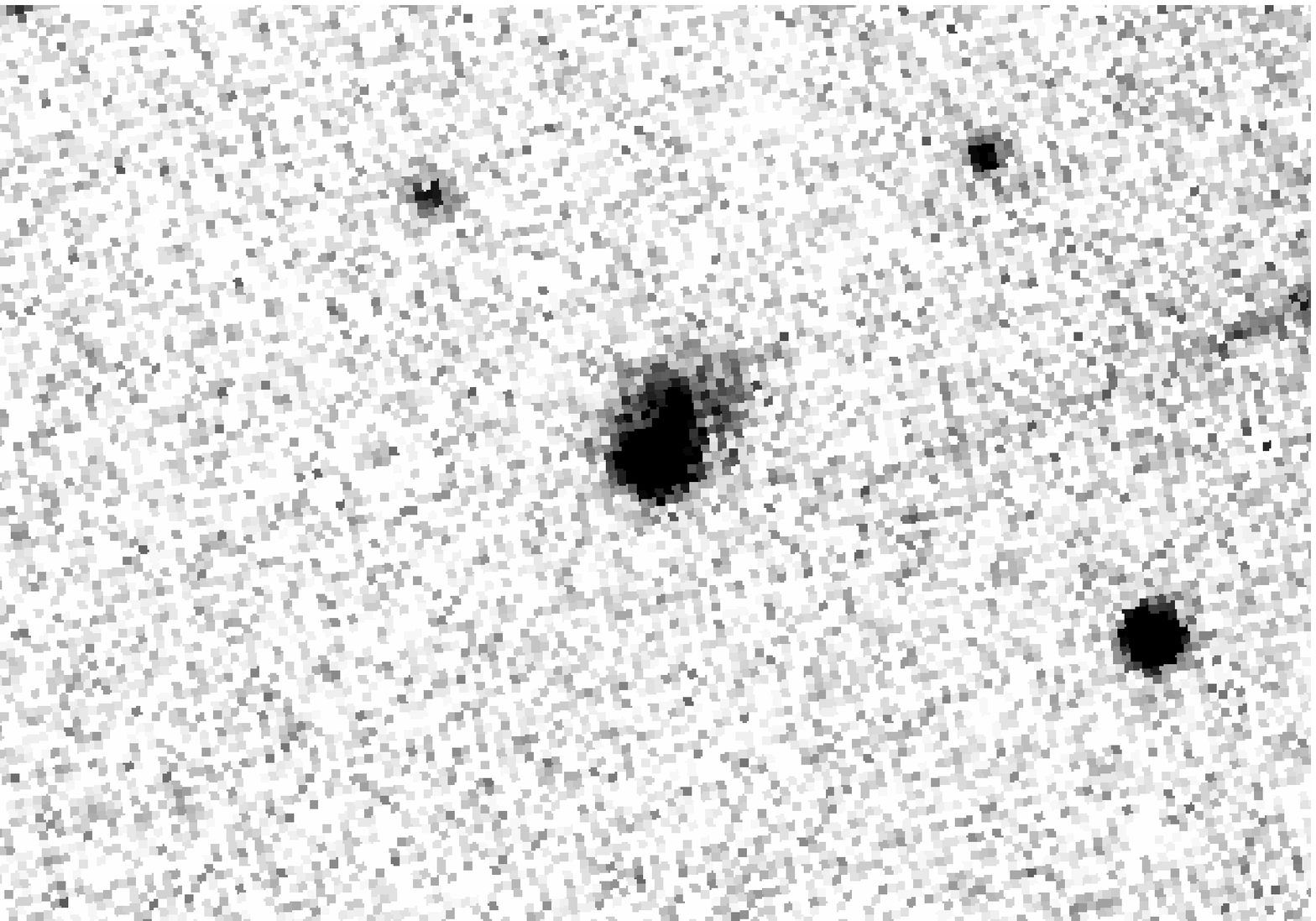}  
\put(3,3){ LAMOST-2}
\end{overpic}
\begin{overpic}[width=0.315\textwidth]{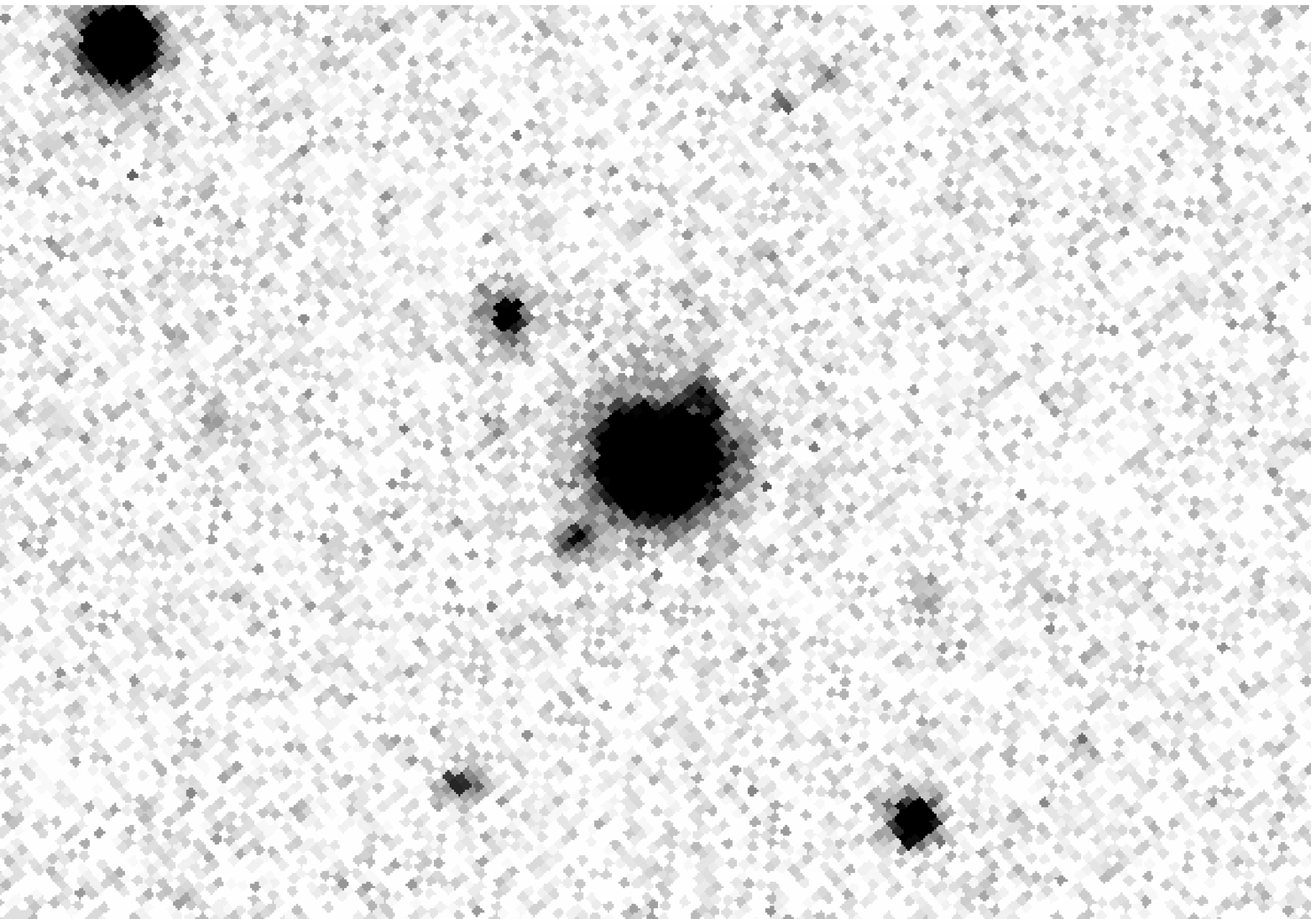} 
\put(3,3){ LAMOST-3}
\end{overpic} \\
\begin{overpic}[width=0.315\textwidth]{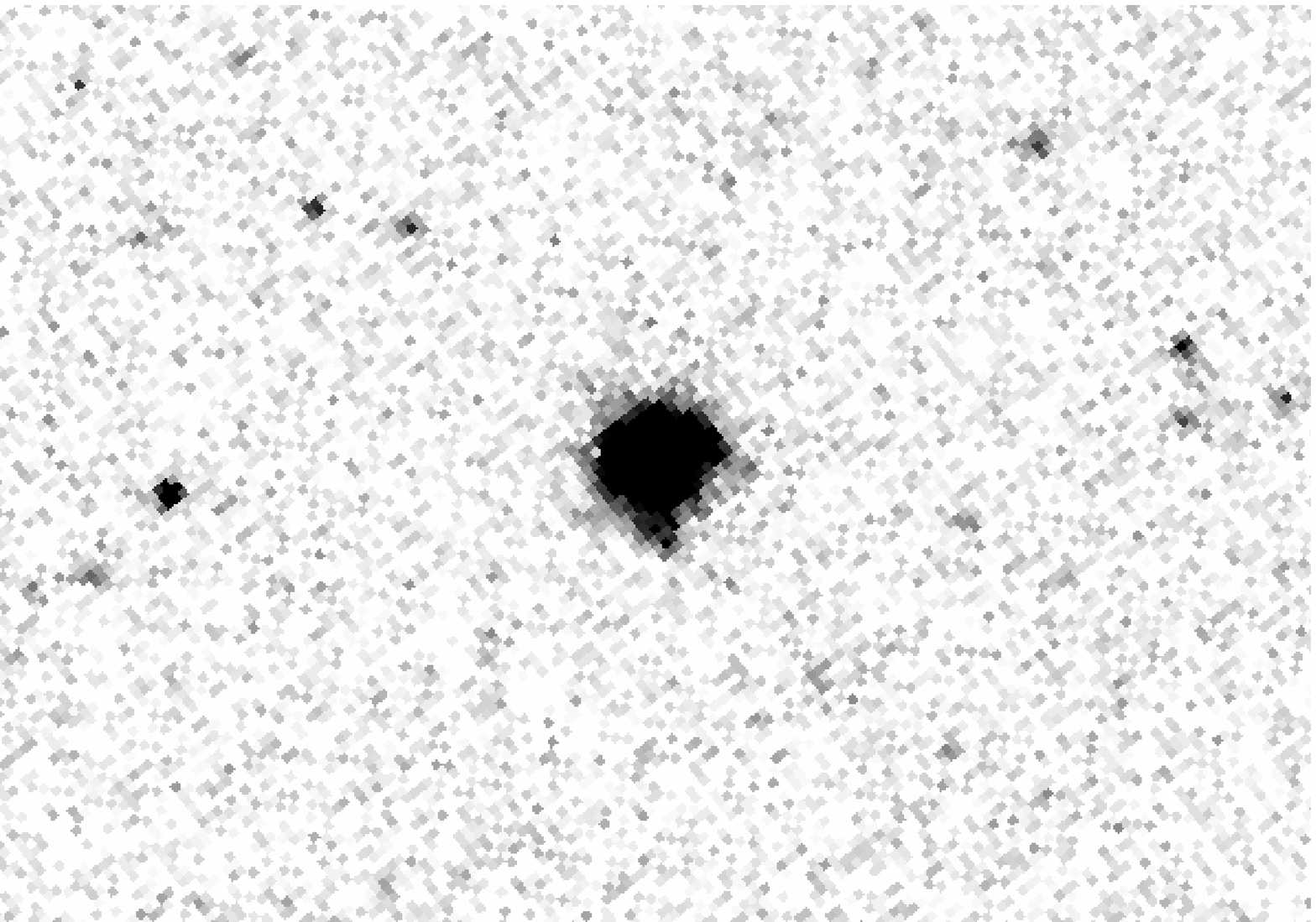} 
\put(3,3){ LAMOST-4}
\end{overpic} 
\begin{overpic}[width=0.315\textwidth]{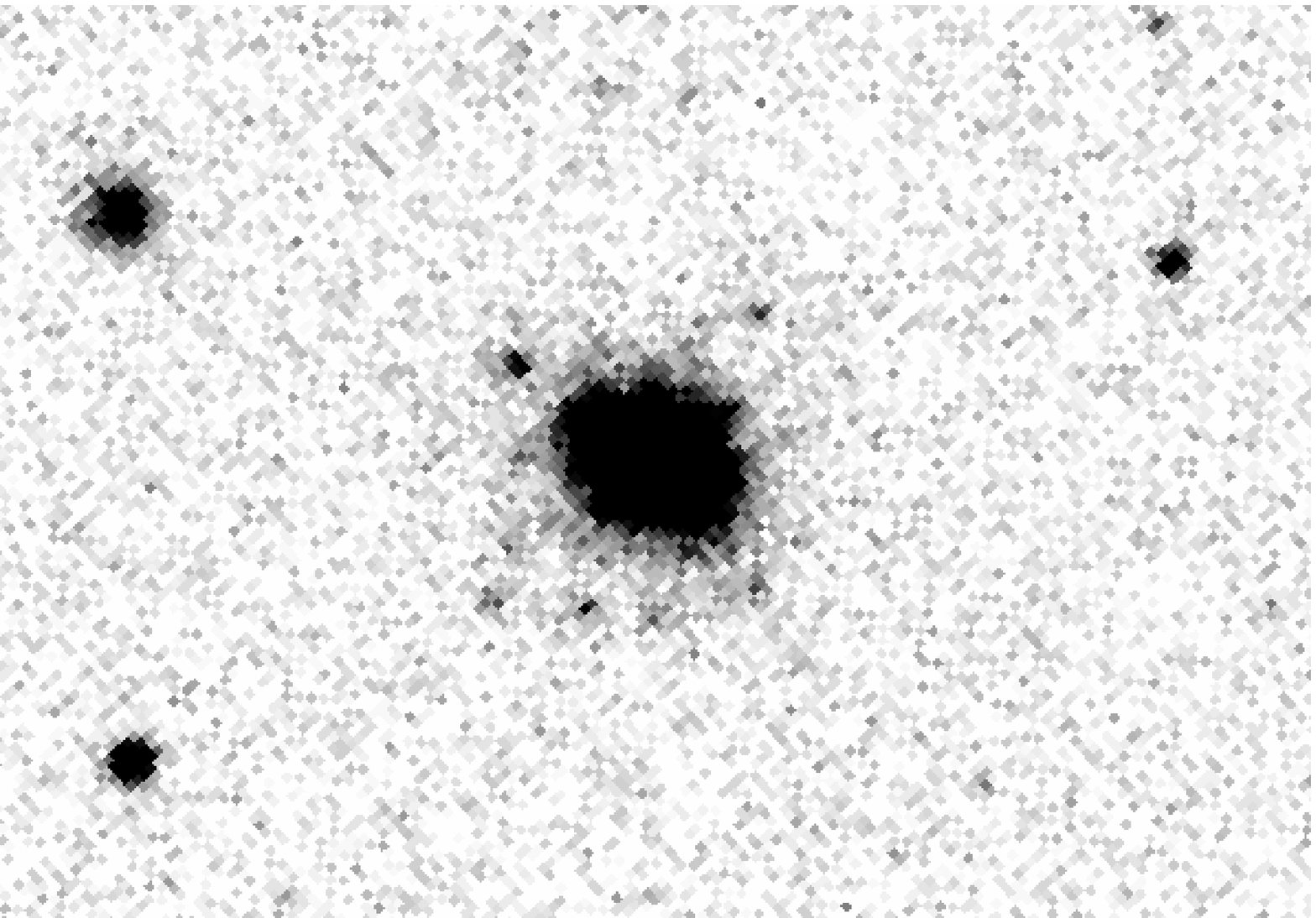} 
\put(3,3){ LAMOST-5}
\end{overpic} 
\caption{Thumbnails of SDSS $r-$band images of the five  objects listed in Table~2. The object names are  labelled.  
Each thumbnail is about 1$' \times 0.7'$ in size, with north at the top and east to the left.} 
\label{gcnim}
\end{figure*}

The positions and radial velocities of the resulted 5 GCs 
from the SDSS Sample are listed in Table~2. 
Two are new discoveries, the other three are recently discovered 
independently  by \citet[LAMOST-4]{diTull2014} 
and \citet[LAMOST-3 and 5]{Huxor2014}.  
LAMOST-3 and LAMOST-5 have been
spectroscopically observed recently by \citet{Velj2014}. The radial velocities derived by \citet{Velj2014}
are consistent with our measurements. For LAMOST-1, 2 and 4, 
no spectroscopic observations have been published yet.
Also listed in Table~2 are their SDSS colors and magnitudes, $g-r$,  $r-i$ and $i$, 
based on the model magnitudes yielded by the SDSS photometric pipeline. The 
differences between the $i$-band PSF and model magnitudes for all objects are also listed in the Table. 
From values of the foreground extinction as given by the extinction map of \citet{Schlafly2014} 
for the M\,31 and M\,33 vicinity fields,
and a distance of 770\,kpc to M31 as given by  \citet{Caldwell2011}, we estimate the absolute $V$-band magnitude of  these 
objects\footnote{For photometric transformation, we use the relation of \citet{Lupton2005},  
$V = g - 0.5784(g - r) - 0.0038$\,mag.}. With $M_V < -6$\,mag, they 
appear to be brighter than  the average of M31 GCs, which 
is probably caused by the relatively bright magnitude limit adopted 
by us when selecting the SDSS Sample of GC candidates.

The LAMOST spectra of the five GCs are plotted in Fig.~\ref{gcsp}. 
Only the spectral range, 4000--5400\,{\AA},  used to calculate the radial velocities is shown  
(Chen et al., in prep.). Also overplotted are the best-fit SSP model spectra.
Most spectra have a good  SNR and the model spectra fit the observations nicely.
The spectrum of  LAMOST-4  has a  relatively low SNR, yet the 
correlation between the observed and model spectra remains to be good. 
The spectra differ  from object to object,  indicating their different properties.
However, they all have a relatively weaker \MgI~
(5176.7\,\AA)~ feature, suggesting that they are all very old GCs. GCs of lower metallicities generally  have stronger 
Balmer lines \citep{Vazdekis2010}. Based on this, it seems that 
 LAMOST-5 is a very metal-poor GC, while LAMOST-1 is metal rich. 
Fig.~\ref{gcnim} displays the SDSS $r$-band images of the five objects. 
They all exhibit morphological appearance unambiguous for typical M31 GCs. 

\begin{table*}
 \centering
  \caption{Positions and properties of newly discovered GC candidates.}
  \begin{tabular}{lcrccccccccc}
  \hline
  \hline
Name & RA & Dec. & $X$  & $Y$ & $R_{\rm p}$ & $g-r$ & $r-i$ & $i_{model}$ & $i_{psf}-i_{model}$ & $V_r$ & [Fe/H]$^a$  \\
          & (deg) & (deg) & (kpc)  & (kpc) & (kpc) & (mag) & (mag) &  (mag) & (mag)  & (\kms)  & (dex) \\
 \hline
   LAMOST-{C01} &    8.60064 &   32.95713 &  -101.96 &   -49.84 &   113.49 &     1.09 &     0.45 &    15.56 &     0.23 &   -224 &  -0.0 \\
   LAMOST-{C02} &   17.44993 &   44.90916 &    80.05 &   -18.98 &    82.27 &     0.34 &     0.06 &    15.36 &     0.30 &   -155 &  -0.9 \\
   LAMOST-{C03} &   12.31295 &   41.51334 &    12.79 &   -10.80 &    16.74 &     0.88 &     0.34 &    15.51 &     0.21 &    -12 &  -0.1 \\
   LAMOST-{C04} &   22.98796 &   30.96388 &   -14.04 &  -189.52 &   190.04 &     1.26 &     0.47 &    14.92 &     0.26 &    -51 &  -0.0 \\
   LAMOST-{C05} &   20.15258 &   36.50384 &    16.66 &  -116.19 &   117.38 &     0.67 &     0.29 &    17.85 &     0.63 &    -60 &  -1.9 \\
   LAMOST-{C06} &    9.77655 &   41.91114 &     1.24 &    12.50 &    12.56 &     1.24 &     0.52 &    15.54 &     0.24 &    -28 &   0.1 \\
   LAMOST-{C07} &   10.13945 &   42.48282 &     9.54 &    14.31 &    17.20 &     0.54 &     0.17 &    14.52 &     0.25 &    -23 &   0.2 \\
   LAMOST-{C08} &   11.51354 &   42.71485 &    20.38 &     5.53 &    21.12 &     0.66 &     0.08 &    14.62 &     0.30 &     -8 &  -0.2 \\
   LAMOST-{C09} &   11.56175 &   42.85695 &    22.17 &     6.35 &    23.06 &     0.52 &     0.31 &    14.83 &     0.33 &    -26 &  -0.2 \\
   LAMOST-{C10} &   14.79318 &   43.93129 &    53.38 &    -8.71 &    54.09 &     0.76 &     0.30 &    15.12 &     0.27 &    -26 &  -0.7 \\
   LAMOST-{C11} &   15.96447 &   45.29663 &    74.50 &    -5.05 &    74.67 &     0.59 &     0.21 &    15.62 &     0.22 &    -39 &  -0.2 \\
   LAMOST-{C12} &   23.75001 &   32.17213 &     3.55 &  -184.21 &   184.24 &     0.71 &     0.32 &    16.97 &     0.47 &    -33 &  -1.3 \\
   LAMOST-{C13} &   20.41774 &   42.53701 &    76.72 &   -61.78 &    98.51 &     0.90 &     0.18 &    14.09 &     0.24 &    -42 &  -0.4 \\
   LAMOST-{C14} &   16.00398 &   40.59559 &    27.55 &   -47.26 &    54.71 &     0.96 &     0.40 &    15.15 &     1.28 &    -61 &  -1.3 \\
   LAMOST-{C15} &   11.48103 &   26.65204 &  -147.20 &  -127.14 &   194.51 &     0.42 &     0.37 &    14.18 &     0.22 &    -49 &  -0.3 \\
   LAMOST-{C16} &   25.11528 &   30.62788 &     0.48 &  -209.17 &   209.17 &     0.90 &     0.45 &    15.34 &     1.43 &      0 &  -0.2 \\
   LAMOST-{C17} &   25.05346 &   25.09957 &   -51.10 &  -259.52 &   264.50 &     0.94 &     0.51 &    15.79 &     1.44 &      8 &   0.0 \\
   LAMOST-{C18} &   23.84216 &   29.55245 &   -20.21 &  -209.28 &   210.25 &     0.90 &     0.43 &    15.70 &     1.29 &    -11 &  -0.1 \\
   LAMOST-{C19} &   26.13748 &   30.74861 &    10.20 &  -215.79 &   216.03 &     0.87 &     0.47 &    15.88 &     1.36 &    -85 &  -0.1 \\
   LAMOST-{C20} &   26.02594 &   30.52029 &     7.16 &  -217.11 &   217.22 &     0.95 &     0.43 &    15.27 &     1.38 &    -94 &  -0.2 \\
   LAMOST-{C21} &   12.47521 &   25.83369 &  -147.98 &  -143.10 &   205.86 &     0.65 &     0.33 &    15.32 &     1.09 &    -50 &  -1.8 \\
   LAMOST-{C22} &   11.73863 &   29.69350 &  -114.12 &  -104.78 &   154.92 &     1.08 &     0.48 &    15.89 &     1.38 &    -26 &  -0.7 \\
   LAMOST-{C23} &   21.04854 &   41.70561 &    73.17 &   -74.07 &   104.12 &     0.91 &     0.47 &    15.66 &     1.40 &    -59 &  -1.5 \\
 \hline
\end{tabular}\\
\begin{flushleft}
$^a$ Metallicities yielded by the  LSP3 which assumes all objects as stars \citep{Xiang2015b}. 
\end{flushleft}
\end{table*}

Table~3 lists the 23 newly discovered candidate clusters 
from the SDSS Sample. They all have radial velocities and spectral types compatible with
being M31 GCs. Their morphologies also look like M31 GCs rather than stars or galaxies. 
Unfortunately,  the SDSS image quality of those objects are not good enough to make a firm conclusion.
To confirm whether they are GCs, images of better  resolution are preferred. 
The radial velocities of these new GC candidates range between $-$224 and 8\,\kms. 
In Table~3 we also list their metallicities yielded by the LSP3 \citep{Xiang2015b} for reference. 
Note that the  LSP3 treats all
objects as stars. If these objects are indeed stars, then the LSP3 results show that most of them are relatively 
metal-rich, with a  metallicity [Fe/H] $\sim$ 0\,dex, suggesting that they are unlikely to be foreground MW halo stars.
This is another indirect indication that they are probably  M31 GCs.
Some of them are quite  extended, such as LAMOST-{C14, C17 and C23}. Thus it is possible that some of  
them could be ultra-compact dwarf galaxies.

\begin{figure}
\centering
   \includegraphics[width=0.68\textwidth]{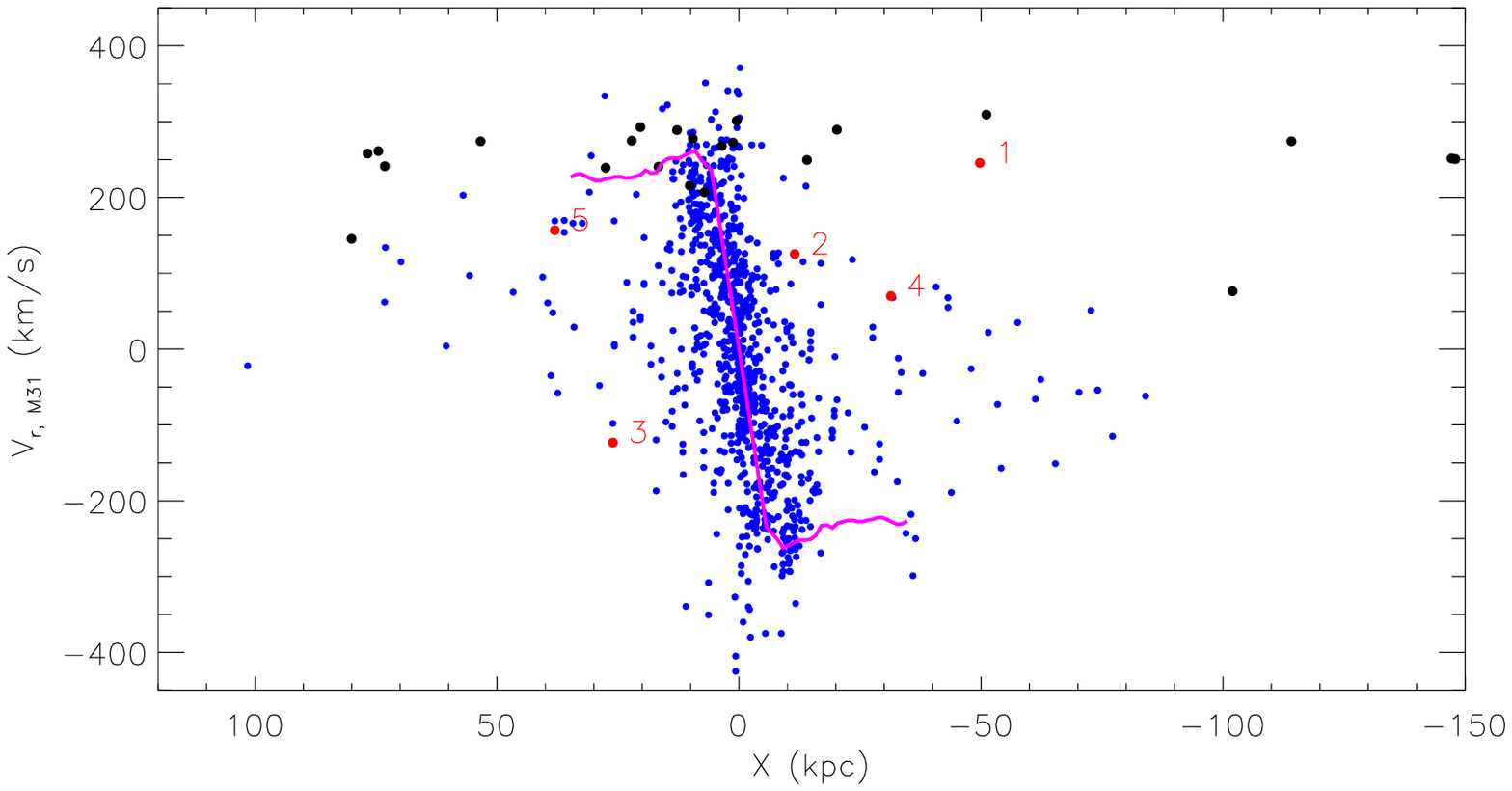}  
   \includegraphics[width=0.68\textwidth]{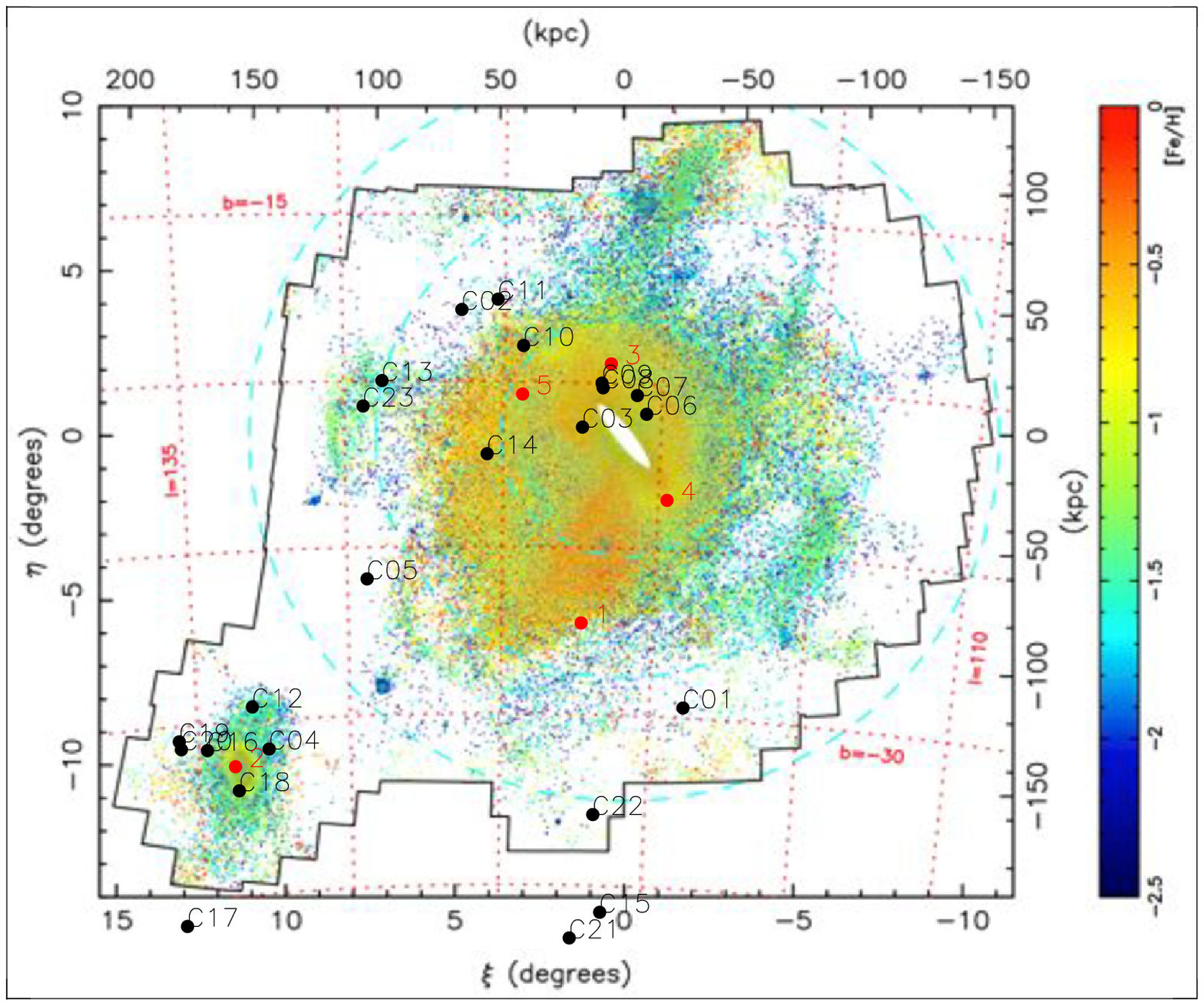}  
\caption{{\it Upper panel}: Radial velocities plotted against the major axis distances
 for previously confirmed M31 GCs (blue dots), 
newly identified  GC candidates (black dots) 
and genuine GCs (red dots). 
The pink line is the \HI~ rotation curve from 
\citet{Carignan2006}. $V_{\rm r,M 31}$ is the line-of-sight velocity in the Andromeda-centric reference system. 
{\it Lower panel}: Spatial distribution of newly identified GCs 
 and candidates, overplotted on a stellar metallicity map from the 
PAndAS survey \citep{Ibata2014}. Red and black dots
represent bona fide GCs and  candidates, respectively.} 
\label{gcpos}
\end{figure}

In Fig.~\ref{gcpos} we present the radial velocity  and spatial distributions of these newly confirmed M31 GCs
and candidates. In the upper panel,  previously known GCs 
from RBC \citep{Galleti2006}, \citet{Caldwell2009} and \citet{Velj2014} 
are also overplotted, along the \HI~ rotation curve from \citet{Carignan2006}. 
The new objects spread over a wide area around M31 and M33,
especially south of M31, where by definition, $X$,  the projected distance along the 
major axis of M31 is negative,  $X<0$\,kpc. Two candidates, LAMOST-{C15} and LAMOST-{C21}, 
have a projected distance $X$ close to $-150$\,kpc. Six objects fall at 
 relatively small projected distances to the center of M31, $13<R_{\rm p} < 30$\,kpc. 
They fit much better the rotation curve than the more remote ones of $R_{\rm p}>30$\,kpc.
Most of the new remote GCs and candidates at large projected distances 
show almost no correlation with the M31 rotation, as for those known M31 halo GCs,
suggesting that they are  possibly accreted ones. 
The newly confirmed GC, LAMOST-1, falls on the Giant  Stream and LAMOST-4 on G1 Clump.
Two candidates, LAMOST-{C13 and C23} locate in the Eastern Cloud. LAMOST-7 and {C11} lie in the North-East 
Structure, while  LAMOST-{C10 and C14} in Stream D. Interestingly, LAMOST-{C14} appears look like an
ultra-compact dwarf galaxy  morphologically, possibly the remnant of an accreted and tidally disrupted dwarf galaxy.
These objects thus provide good opportunity study the assemblage history of M31 halo (Chen et al. in prep.).

LAMOST-2 falls in the halo region of M33  (see the lower panel of Fig.~\ref{gcpos}), 
together with 6 GC candidates.  These objects have radial velocities ranging from $-175$ to
$0$\,\kms, close to the M33 system radial velocity of  $-179.2\,\rm km\,s^{-1}$ \citep{McConnachie2012}. 
In particular, LAMOST-2, the GC  closest
to M33, has a radial velocity of  $-175\,$\kms. 
They could thus be GCs belonging to M33.  
LAMOST-1 is the most distant  
confirmed GC from M31 in our sample, 
except for those possibly belonging to M33. It has a  projected distance from M31 of 
$R_{\rm p} = 78$\,kpc. LAMOST-{C15, C17 and C21}  have very large distances from  both M31 and M33.
They are possibly  intergalactic GCs.


\subsection{Updates to the Literature Sample}

\begin{figure*}
\centering
   \includegraphics[width=0.98\textwidth]{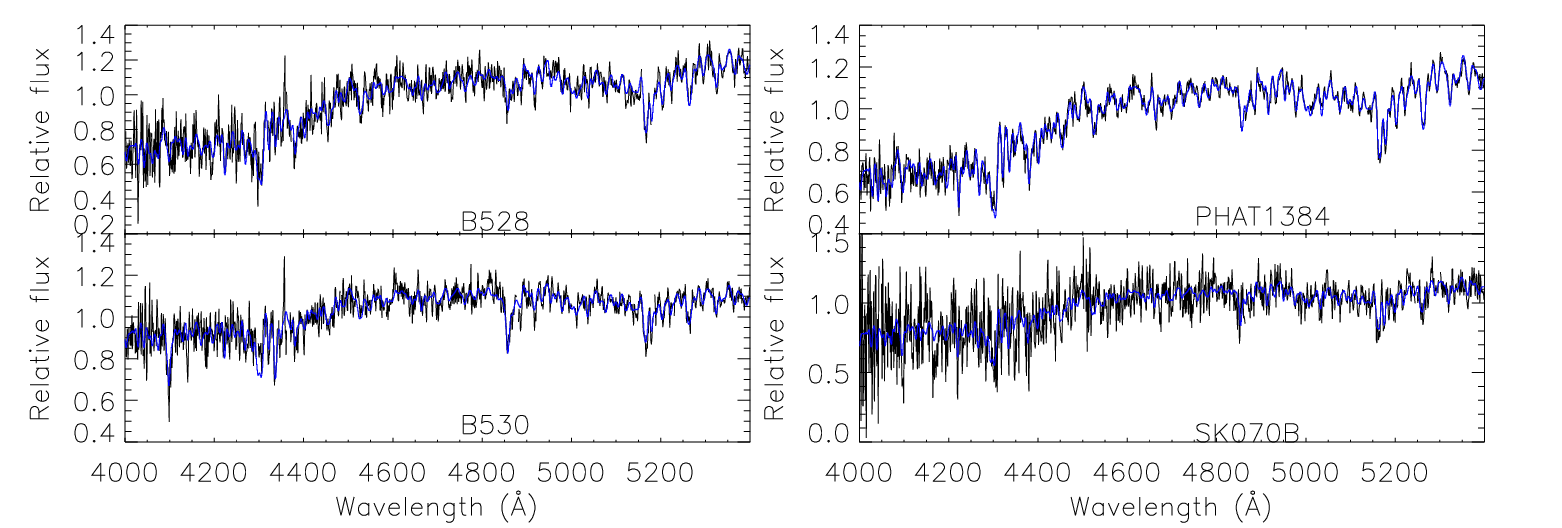}  
\caption{Same as Fig.~\ref{gcsp} but for newly confirmed GCs from the Literature Sample.} 
\label{rbcsp2}
\end{figure*}

\begin{table*}
 \caption{Updates to classification of objects from the RBC V5 and from \citet{Johnson2012}.}
\begin{tabular}{lcrcccccccc}
  \hline
  \hline
Name & RA & Dec. & $X$  & $Y$ & $R_{\rm p}$ & New $V_r $ & RBC $V_r$ & Previous & New  \\
          & (deg) & (deg) & (kpc)  & (kpc) & (kpc) & (\kms) & (\kms) & classification & classification  \\
 \hline
\textbf{RBC V5}\\
         B530 &   10.70963 &   41.42019 &     1.75 &     1.05 &     2.05 &   -239 &  99999 &  2 &  1 \\
         B528 &   10.79958 &   41.39664 &     2.06 &     0.14 &     2.07 &   -197 &  99999 &  2 &  1 \\
       SK070B &   10.28720 &   41.16546 &    -3.57 &     2.32 &     4.25 &   -463 &  99999 &  2 &  1 \\
        B233D &   10.92214 &   39.61274 &   -16.02 &   -15.64 &    22.39 &  99999 &  99999 &  2 &  4 \\
        B334D &   12.22861 &   39.59899 &    -7.73 &   -26.33 &    27.44 &  99999 &  99999 &  2 &  4 \\
        B186D &   10.00940 &   39.38665 &   -24.23 &   -10.03 &    26.22 &  99999 &  99999 &  2 &  4 \\
         B412 &    8.73034 &   41.54069 &    -9.05 &    17.87 &    20.03 &  99999 &  99999 &  2 &  4 \\
         B504 &   12.18817 &   40.14620 &    -2.28 &   -21.38 &    21.50 &  99999 &  99999 &  2 &  4 \\
       SK190B &   11.25680 &   40.40317 &    -5.55 &   -11.77 &    13.01 &  99999 &  99999 &  2 &  4 \\
       SK048C &    9.40504 &   40.09456 &   -20.46 &     0.71 &    20.47 &  99999 &  99999 &  2 &  4 \\
         B411 &    8.62837 &   41.56225 &    -9.43 &    18.87 &    21.09 &  99999 &  99999 &  2 &  4 \\
        B339D &   12.32291 &   40.75195 &     4.91 &   -17.32 &    18.01 &  99999 &  99999 &  2 &  4 \\
         B413 &    8.80411 &   41.48547 &    -9.20 &    16.83 &    19.18 &  99999 &  99999 &  2 &  4 \\
       SK102B &   10.59450 &   40.37323 &   -10.06 &    -6.69 &    12.07 &  99999 &  99999 &  2 &  4 \\
         WH23 &   11.25420 &   41.51660 &     6.16 &    -2.46 &     6.64 &   -163 &   -159 &  6 &  5 \\
       SK026B &    9.70296 &   40.52194 &   -14.04 &     1.75 &    14.15 &   -491 &  99999 &  6 &  5 \\
         SH06 &    9.82938 &   40.36611 &   -14.92 &    -0.54 &    14.93 &  -1047 &  99999 &  2 &  5 \\

\textbf{PHAT}\\
         PHAT1384 &   10.67271 &   41.31956 &     0.46 &     0.51 &     0.69 &   -284 &  99999 &  2 &  1 \\
         PHAT273 &   11.24446 &   41.92677 &    10.42 &     1.04 &    10.48 &   -158 &  99999 &  2 &  5 \\
         PHAT1495 &   11.22254 &   42.04267 &    11.51 &     2.18 &    11.71 &    99999 &  99999 &  2 &  5 \\
         
\hline
\end{tabular}\\
  \label{ta4}
\end{table*}

Except for candidates from the SDSS Sample, there are 355 known clusters and candidates
from the Literature Sample. They mostly fall near the disk of M31, 
within a  projected distance $R_{\rm p} < 30$\,kpc. 
Most of the  sources are from  the RBC. 
A couple of objects are recent discoveries from the 
Panchromatic Hubble Andromeda Treasury (PHAT) survey \citep{Johnson2012}.  
\citet{Johnson2012} have recently published a catalog containing  
601 `confirmed' clusters (noted as `c') and 237 `probable' candidates (noted as `p'). 
These clusters seem to be less massive than the typical GCs discussed in the current work \citep{Johnson2012}.
In our Literature Sample, there are 350 objects from the RBC and 5 objects from 
\citet{Johnson2012}, respectively. Given their closeness to the M31 disk, the 
SDSS images are not of much help for their identifications. We classify them  based on the LAMOST spectra only. 
Some of  the targets have  spectra  characteristic of, e.g.  
M-type stars, galaxies and \HII~ regions, and are classified accordingly. 
If the spectrum has the characteristics of a GC and yields a
radial velocity smaller than $-150\,$\kms, the object is classified as a bona fide GC

296 objects with an original RBC class of 1 or 8  are classified as genuine clusters. 
These objects are listed in Table~6. The radial velocities 
newly derived from our spectra, as well as original values  from the RBC are  
listed in the Table. The spectra are all consistent of being 
GCs. The  radial velocities ranging between $-726$ and $39\,$\kms, which are inside the 
radial velocity range for possible M31 objects.
The radial velocities are in good agreement with those from the RBC, with an average difference and scatter
of $V_r-V_{r,{\rm RBC}}=-11 \pm 22$\,\kms. 
38 objects are classified as GC candidates in the
RBC (class 2). We find that 3 have radial velocities $V_r < -150\,\rm km\,s^{-1}$. They are classified 
as genuine GCs (class 1) in the current work. 11 candidates have large red-shifts and they are classified as 
background galaxies (class 4). One target shows strong narrow emission lines classic for \HII~ regions 
(see the third example spectrum in Fig.~4). Its classification is modified 
to \HII~ regions (class 5). The remaining  candidates have 
radial velocities $V_r > -150$\,\kms~ and show no obvious spectral features to help classify them. 
They could be GCs or foreground MW stars. We retain their classifications as candidate GCs (class 2).
Morphological information could be quite helpful for more secure classifications.
Two objects originally classified as stars in the RBC (class 6) show spectral features 
of \HII~ regions and they are reclassified accordingly (class 5). 
LAMOST spectra of the newly confirmed GCs from the Literature Sample are plotted in Fig.~\ref{rbcsp2} and 
updates to the classifications of objects from the RBC V5 are listed in Table~4.
The newly derived radial velocities (for background galaxies, 
their radial velocities are noted as ``99999'') ,
their original values from the RBC, and the new classifications are listed in the Table.

In Table~4 we also list 3 targets from \citet{Johnson2012} classified by them as 
possible clusters. We find that one of them (PHAT1384) has a radial 
velocity of $-284\,$\kms, and is reclassified as 
bona fide cluster in the current work. The LAMOST spectrum of PHAT1384 is 
plotted in Fig.~\ref{rbcsp2}.
The remaining two candidates  show spectral features characteristic of 
 \HII~ regions and are reclassified accordingly.
Two objects from \citet{Johnson2012}, classified as clusters  by them,
have spectra compatible of being GCs and have radial velocities of $-81$ (PHAT224) and $-119\,\rm km\,s^{-1}$
(PHAT1148). These two objects are listed in Table~6 as known clusters.

\section{Summary}

\begin{table}
\centering
 \caption{Summary of the catalogs.}
\begin{tabular}{llc}
  \hline
  \hline
 & Type & Number \\
 \hline
SDSS sample & New bona fide GCs & 5 \\
 (Tables~2 and 3)  & New candidate GCs  & 23\\
                     & Background galaxies & 218\\
                     & Foreground stars & 22 \\
                     & Quasars & 8\\
 & Unknown objects  & 277\\
 \hline
Literature sample & Newly confirmed GCs & 4 \\
(Tables 3, 4 and 5)  & Previous known GCs & {298} \\
   &   GC candidates &  26\\
   & \HII~ regions & 11\\
   & Galaxies & 16\\
\hline
\end{tabular}\\
  \label{ta5}
\end{table}

We present a catalog of {908} targets observed as M31 GC candidates with LAMOST 
from September, 2011 to June, 2014. A summary of the catalog is presented in Table~5.
We have searched for distant clusters amongst targets selected 
from the SDSS photometry of non-stellar objects in the outer halos of M31 and M33. By combining 
information from the LAMOST spectra and morphological information from 
 the SDSS images, 
we identified 5 bona fide GCs and 23 candidates, amongst them 25 are 
newly discoveries. One of the confirmed GC, LAMOST-2, falls in the halo of M33. Its
radial velocity is compatible for being a  M33 GC. 
The other  newly discovered  bona 
fide GC (LAMOST-1) falls on the Giant Stream with a projected distance of  78\,kpc from M31.
The newly identified GC candidates have a maximum projected distance  of  265\,kpc (LAMOST-{C17})
from M31. 
In addition, 218 background galaxies, 8 quasars, 1 
white-dwarf-and-main-sequence-star binary,  and 21 stars are identified amongst our targets.
There are 277 objects exhibiting ambiguous  radial velocities and spectral types 
and cannot be reliably classified as either 
M31 GCs or foreground MW stars. Images of better resolution 
than available from the SDSS may help classfy these objects. 

We have also observed some known GCs and GC candidates selected from the literature. Lists of these
objects are provided, including {298} previously known GCs and 32 candidates, 
containing newly derived radial velocities and updates to the classifications. We have 
also identified  4 candidates as bona fide M31 GCs based on the newly deduced radial velocities.
The LAMOST observations of GCs and GC candidates  
provide us a great opportunity to study the GC system of M31.
In the current work,  {307} confirmed 
GCs observed by LAMOST are cataloged.
The observations, combined with photometric data,  yield  information 
of the  kinematics, chemistry and age.
A full analysis of this data set will be presented in a separate paper.

\begin{table*}
\scriptsize
 \caption{Known GCs from the RBC V5 and from \citet{Johnson2012} observed by LAMOST.}
\begin{tabular}{lcccccclcccccc}
  \hline
  \hline
Name & RA & Dec. & $R_{\rm p}$ & New $V_r $ & RBC $V_r$ & &Name & RA & Dec. & $R_{\rm p}$ & New $V_r $ & RBC $V_r$  \\ 
          & (deg) & (deg) & (kpc) & (\kms) & (\kms) &           &         & (deg) & (deg) & (kpc) & (\kms) & (\kms)          \\
 \hline
         NB21 &   10.65825 &   41.26636 &    0.85 &   -350 &   -773 &  &          B131 &   10.71167 &   41.28542 &    0.50 &   -328 &   -424 \\
         B124 &   10.67261 &   41.25663 &    0.21 &   -255 &     70 &  &          B358 &   10.82438 &   39.82032 &   60.10 &   -315 &   -315 \\
         B167 &   10.83802 &   41.23563 &    6.69 &   -368 &   -231 &  &          B338 &   10.24530 &   40.59659 &   13.36 &   -277 &   -274 \\
         B115 &   10.64337 &   41.23387 &    0.65 &   -467 &   -549 &  &          B086 &   10.57773 &   41.23389 &    2.70 &   -272 &   -189 \\
         B225 &   11.12316 &   41.35993 &   12.68 &   -169 &   -165 &  &          B148 &   10.76607 &   41.30136 &    1.90 &   -299 &   -290 \\
         B405 &   12.41585 &   41.59160 &   50.66 &   -170 &   -162 &  &          VDB0 &   10.12250 &   40.60414 &   11.41 &   -594 &   -567 \\
         B019 &   10.21887 &   41.31483 &   18.34 &   -232 &   -224 &  &          B012 &   10.13525 &   41.36226 &   23.01 &   -367 &   -363 \\
         B240 &   11.35437 &   41.10616 &   29.78 &    -63 &    -56 &  &          B119 &   10.65043 &   41.29322 &    2.09 &   -316 &   -310 \\
         B106 &   10.62934 &   41.20513 &    1.10 &   -252 &   -102 &  &          BH05 &   10.12714 &   40.75814 &    8.96 &   -600 &   -570 \\
         B178 &   10.87825 &   41.35459 &    4.26 &   -164 &   -138 &  &         MCGC8 &   13.60396 &   39.71554 &  161.44 &   -380 &   -381 \\
         B127 &   10.68541 &   41.24484 &    0.96 &   -417 &   -470 &  &          B152 &   10.79167 &   41.30447 &    2.70 &   -185 &   -129 \\
         B384 &   11.59138 &   40.28322 &   68.85 &   -366 &   -364 &  &          B058 &   10.47083 &   40.78602 &   12.02 &   -225 &   -224 \\
         B028 &   10.31871 &   40.98422 &    5.88 &   -449 &   -434 &  &          B114 &   10.64292 &   41.21247 &    1.05 &   -374 &   -271 \\
         B077 &   10.54640 &   41.12612 &    2.40 &   -518 &   -681 &  &          B103 &   10.62390 &   41.29928 &    3.26 &   -365 &   -334 \\
         B218 &   11.05968 &   41.32206 &   11.66 &   -225 &   -220 &  &          B027 &   10.31055 &   40.93081 &    5.98 &   -302 &   -264 \\
         B020 &   10.23026 &   41.69037 &   31.54 &   -358 &   -351 &  &          B110 &   10.63793 &   41.05787 &    6.62 &   -244 &   -247 \\
         B123 &   10.66941 &   41.17595 &    3.09 &   -385 &   -368 &  &          B312 &    9.91738 &   40.95068 &   17.65 &   -188 &   -174 \\
         B144 &   10.74942 &   41.26829 &    2.36 &   -352 &   -140 &  &          B158 &   10.80996 &   41.12253 &    9.87 &   -193 &   -187 \\
         B125 &   10.67612 &   41.09197 &    6.51 &   -607 &   -641 &  &          B344 &   10.51239 &   41.86726 &   28.53 &   -256 &   -252 \\
         B179 &   10.87960 &   41.30407 &    5.83 &   -171 &   -139 &  &          B112 &   10.63853 &   41.29511 &    2.58 &   -303 &   -237 \\
         B472 &   10.95173 &   41.44809 &    4.55 &   -126 &   -106 &  &          B163 &   10.82346 &   41.46252 &    3.66 &   -173 &   -174 \\
         B143 &   10.74850 &   41.32206 &    1.01 &   -182 &   -128 &  &          B205 &   10.99239 &   41.41065 &    6.60 &   -374 &   -373 \\
         B171 &   10.85665 &   41.26032 &    6.48 &   -214 &   -273 &  &          B230 &   11.14660 &   40.95342 &   28.01 &   -586 &   -597 \\
         B213 &   11.01463 &   41.51075 &    5.36 &   -461 &   -545 &  &          B099 &   10.61495 &   41.16747 &    1.97 &   -334 &   -200 \\
         B206 &   10.99424 &   41.50502 &    4.95 &   -193 &   -192 &  &          B107 &   10.63022 &   41.32748 &    4.08 &   -319 &   -337 \\
         B311 &    9.89052 &   40.52075 &   12.93 &   -526 &   -469 &  &          G002 &    8.39076 &   39.52187 &   38.43 &   -357 &   -313 \\
         B073 &   10.53048 &   40.98925 &    6.21 &   -505 &   -481 &  &          B324 &   10.08529 &   41.68036 &   36.26 &   -254 &   -299 \\
         B212 &   11.01281 &   41.08231 &   18.50 &   -406 &   -402 &  &          B343 &   10.42958 &   40.20623 &   32.56 &   -361 &   -360 \\
         B064 &   10.50799 &   41.18541 &    3.75 &   -351 &   -307 &  &          B187 &   10.91093 &   41.49634 &    3.83 &   -109 &   -130 \\
         B101 &   10.62093 &   41.13767 &    3.14 &   -401 &   -489 &  &          B403 &   12.32344 &   41.58564 &   47.62 &   -276 &   -358 \\
         B059 &   10.47539 &   41.18354 &    4.81 &   -346 &   -332 &  &          B153 &   10.79421 &   41.24760 &    4.69 &   -262 &   -231 \\
         B063 &   10.50360 &   41.48602 &   14.41 &   -311 &   -302 &  &          B085 &   10.57599 &   40.66588 &   19.63 &   -433 &   -526 \\
         B005 &   10.08462 &   40.73287 &    9.58 &   -299 &   -265 &  &          B017 &   10.20303 &   41.20197 &   15.08 &   -544 &   -514 \\
        B344D &   12.46693 &   41.61080 &   51.81 &   -168 &     21 &  &          B193 &   10.93962 &   41.61601 &    6.48 &    -79 &    -59 \\
         EXT8 &   13.31045 &   41.55686 &   83.14 &   -214 &   -154 &  &          B006 &   10.11031 &   41.45740 &   27.28 &   -244 &   -237 \\
         B386 &   11.61255 &   42.03132 &   14.46 &   -403 &   -383 &  &          B088 &   10.58777 &   41.53729 &   13.46 &   -512 &   -478 \\
         B357 &   10.80513 &   40.18236 &   45.59 &   -348 &   -285 &  &          B190 &   10.93074 &   41.56832 &    5.25 &   -119 &    -86 \\
         B289 &    8.58700 &   41.79753 &   94.00 &   -174 &   -181 &  &          B345 &   10.55888 &   40.29345 &   33.28 &   -361 &   -293 \\
         G327 &   11.70622 &   42.74636 &   29.10 &   -285 &   -251 &  &          B081 &   10.55664 &   40.81083 &   13.53 &   -411 &   -430 \\
         B361 &   10.98787 &   40.23375 &   49.79 &   -353 &   -330 &  &          B034 &   10.36715 &   40.89714 &    6.38 &   -552 &   -539 \\
         B076 &   10.54264 &   41.08946 &    3.20 &   -533 &   -520 &  &          B105 &   10.62810 &   41.50759 &   10.98 &   -264 &   -238 \\
         B302 &    9.63964 &   41.34790 &   40.39 &   -428 &   -371 &  &          B381 &   11.52728 &   41.34969 &   27.40 &    -78 &    -86 \\
         B066 &   10.51287 &   40.74642 &   14.67 &   -438 &   -389 &  &          B234 &   11.19334 &   41.48827 &   11.27 &   -204 &   -207 \\
         B350 &   10.61847 &   40.41423 &   30.58 &   -430 &   -467 &  &          B400 &   12.00599 &   42.42583 &   20.63 &   -262 &   -258 \\
         B292 &    9.06941 &   40.97404 &   48.75 &   -374 &   -307 &  &          B201 &   10.97016 &   41.16611 &   13.92 &   -725 &   -706 \\
         B397 &   11.86348 &   41.20290 &   44.48 &   -125 &   -228 &  &          B373 &   11.42436 &   41.75929 &   12.55 &   -229 &   -219 \\
         B233 &   11.17549 &   41.73180 &    7.94 &    -79 &    -74 &  &          B383 &   11.54978 &   41.32822 &   28.93 &   -234 &   -253 \\
         B156 &   10.80719 &   41.02159 &   13.59 &   -384 &   -400 &  &          MGC1 &   12.67691 &   32.91633 &  391.24 &   -355 &   -355 \\
         B039 &   10.40779 &   41.34716 &   12.69 &   -253 &   -243 &  &          B130 &   10.70356 &   41.49794 &    8.15 &    -53 &    -25 \\
         B223 &   11.11271 &   41.57699 &    6.99 &    -74 &   -101 &  &          B174 &   10.87623 &   41.64896 &    8.91 &   -489 &   -478 \\
         B161 &   10.81421 &   41.19027 &    7.49 &   -436 &   -413 &  &          B082 &   10.56601 &   41.02063 &    5.97 &   -383 &   -366 \\
         B065 &   10.50807 &   40.67026 &   17.40 &   -404 &   -382 &  &          B339 &   10.25296 &   39.93169 &   37.59 &   -242 &   -188 \\
         B293 &    9.08691 &   40.89363 &   45.60 &   -526 &   -424 &  &        SK007A &    9.93363 &   40.85526 &   14.72 &   -446 &   -390 \\
         B317 &    9.98030 &   41.79614 &   44.20 &   -202 &   -143 &  &          B060 &   10.48750 &   41.08735 &    3.17 &   -537 &   -484 \\
         B098 &   10.61408 &   40.99333 &    8.35 &   -314 &   -268 &  &          B094 &   10.60434 &   40.95489 &    9.51 &   -564 &   -565 \\
         B091 &   10.59046 &   41.36817 &    6.98 &   -302 &   -290 &  &          B188 &   10.92297 &   41.40721 &    4.45 &   -185 &   -184 \\
         B365 &   11.15185 &   42.28902 &   25.28 &    -76 &    -67 &  &          B025 &   10.30230 &   41.00781 &    6.51 &   -249 &   -206 \\
         B313 &    9.93587 &   40.88195 &   15.26 &   -439 &   -440 &  &          B352 &   10.65911 &   42.03696 &   30.20 &   -298 &   -164 \\
         B204 &   10.98510 &   41.36747 &    7.57 &   -340 &   -352 &  &          B096 &   10.60863 &   41.32072 &    4.58 &   -319 &   -260 \\
         B157 &   10.80821 &   41.18880 &    7.33 &   -266 &      4 &  &          B180 &   10.88221 &   41.12952 &   12.14 &   -213 &   -202 \\
         B038 &   10.39983 &   41.32077 &   12.04 &   -191 &   -175 &  &          B042 &   10.42367 &   41.12394 &    5.04 &   -313 &   -338 \\
         V014 &   10.30755 &   40.56607 &   15.77 &   -467 &   -450 &  &          B080 &   10.55160 &   41.31685 &    6.47 &   -298 &   -255 \\
        B087D &   10.74546 &   41.15242 &    6.51 &   -452 &   -661 &  &          B315 &    9.95223 &   40.52513 &   12.51 &   -560 &   -437 \\
         B399 &   11.99811 &   41.59126 &   36.05 &   -426 &   -437 &  &          B401 &   12.03546 &   41.67830 &   34.60 &   -352 &   -333 \\
         B134 &   10.71522 &   41.23433 &    2.37 &   -274 &   -369 &  &          B291 &    9.02071 &   42.03592 &   86.83 &   -220 &   -215 \\
         B323 &   10.07620 &   40.54580 &   12.43 &   -525 &   -500 &  &          B001 &    9.96253 &   40.96963 &   16.60 &   -222 &   -179 \\
         B306 &    9.78627 &   40.57250 &   14.58 &   -440 &   -424 &  &          G260 &   11.00352 &   42.58008 &   40.40 &   -213 &     16 \\
         B407 &   12.54149 &   41.68366 &   52.01 &   -315 &   -338 &  &         B020D &   10.32182 &   41.13588 &    8.78 &   -479 &   -526 \\
         B194 &   10.93828 &   41.10239 &   15.12 &   -401 &   -354 &  &          B141 &   10.74703 &   41.54651 &    8.66 &   -187 &   -180 \\
         B321 &   10.06409 &   40.46276 &   14.45 &   -516 &   -519 &  &          B184 &   10.90626 &   41.60961 &    6.87 &   -160 &   -152 \\
         B208 &   11.00032 &   41.38654 &    7.54 &   -218 &   -222 &  &          B050 &   10.44285 &   41.53846 &   18.48 &   -118 &   -114 \\
             \hline
\end{tabular}\\
\end{table*} 

\addtocounter{table}{-1}
\begin{table*}
\scriptsize
 \caption{Continued.}
\begin{tabular}{lcccccclcccccc}
  \hline
  \hline
Name & RA & Dec. & $R_{\rm p}$ & New $V_r $ & RBC $V_r$ & &Name & RA & Dec. & $R_{\rm p}$ & New $V_r $ & RBC $V_r$  \\ 
          & (deg) & (deg) & (kpc) & (\kms) & (\kms) &           &         & (deg) & (deg) & (kpc) & (\kms) & (\kms)          \\
 \hline
	B235 &   11.24136 &   41.48998 &   12.85 &    -97 &    -98 &  &          B051 &   10.44454 &   41.42199 &   14.10 &   -258 &   -270 \\
         B486 &   11.53571 &   40.96769 &   41.28 &   -204 &   -152 &  &          B202 &   10.97784 &   41.00895 &   20.00 &   -246 &   -343 \\
         B257 &   10.51371 &   40.97052 &    6.48 &   -486 &   -476 &  &          B004 &   10.07462 &   41.37787 &   25.74 &   -381 &   -369 \\
         B337 &   10.20202 &   42.20304 &   51.68 &   -294 &     50 &  &          B135 &   10.71653 &   41.51895 &    8.55 &   -372 &   -371 \\
         B217 &   11.04416 &   41.39752 &    8.73 &    -52 &    -25 &  &          B189 &   10.92663 &   41.58984 &    5.89 &   -143 &   -148 \\
         B176 &   10.87690 &   40.81964 &   23.65 &   -536 &   -521 &  &          B310 &    9.85724 &   41.39256 &   34.08 &   -256 &   -206 \\
         B319 &   10.01279 &   40.56620 &   11.78 &   -565 &   -535 &  &          B370 &   11.31000 &   41.96132 &   11.77 &   -374 &   -352 \\
        B150D &    9.24949 &   41.42505 &   57.16 &   -269 &  99999 &  &          B382 &   11.54312 &   41.62790 &   19.21 &   -272 &   -302 \\
         B216 &   11.03659 &   41.63217 &    6.11 &    -80 &    -93 &  &         B256D &   11.24469 &   41.91020 &   11.01 &    -86 &    -74 \\
         G085 &   10.30333 &   40.57144 &   15.48 &   -468 &   -444 &  &          B031 &   10.33716 &   40.98449 &    5.52 &   -377 &   -400 \\
         B008 &   10.12613 &   41.26907 &   20.11 &   -330 &   -319 &  &          B255 &   10.50003 &   40.80945 &   11.95 &   -438 &   -433 \\
         B097 &   10.61443 &   41.42560 &    8.32 &   -290 &   -283 &  &          B083 &   10.56852 &   41.75572 &   22.42 &   -391 &   -347 \\
         B185 &   10.90534 &   41.24543 &    8.75 &   -198 &   -163 &  &          B036 &   10.38677 &   41.43478 &   16.62 &   -510 &   -341 \\
         B314 &    9.93581 &   40.23553 &   19.14 &   -490 &   -485 &  &          B018 &   10.20583 &   40.69220 &    9.98 &   -593 &   -585 \\
         B393 &   11.75501 &   41.40185 &   33.72 &   -395 &   -331 &  &          NB89 &   10.68658 &   41.24561 &    0.97 &   -280 &   -332 \\
         B391 &   11.74205 &   41.56579 &   27.91 &   -320 &   -325 &  &          B049 &   10.43987 &   40.83194 &    9.59 &   -478 &   -481 \\
         B325 &   10.09632 &   40.51310 &   13.47 &   -636 &   -560 &  &          B147 &   10.76375 &   41.35604 &    1.47 &   -216 &    -51 \\
         B322 &   10.07186 &   40.65123 &   10.40 &   -594 &   -581 &  &          B029 &   10.32433 &   41.00640 &    5.93 &   -516 &   -374 \\
         B111 &   10.63826 &   41.00732 &    8.55 &   -444 &   -414 &  &          B022 &   10.24616 &   41.41172 &   20.79 &   -463 &   -407 \\
         B003 &   10.03917 &   41.18478 &   20.42 &   -381 &   -351 &  &           H19 &   11.06200 &   38.42839 &  121.52 &   -285 &  99999 \\
         B011 &   10.13282 &   41.65474 &   33.64 &   -249 &   -178 &  &          B295 &    9.19471 &   40.32842 &   27.21 &   -418 &   -408 \\
         B448 &   10.15211 &   40.67088 &   10.10 &   -554 &   -552 &  &          G268 &   11.04174 &   42.78273 &   47.01 &   -296 &   -321 \\
         B214 &   11.01641 &   41.43850 &    6.69 &   -183 &   -258 &  &          B301 &    9.58998 &   40.06029 &   20.26 &   -389 &   -382 \\
         B349 &   10.60053 &   40.62886 &   21.81 &   -407 &   -406 &  &          B222 &   11.10563 &   41.23665 &   16.23 &   -324 &   -303 \\
       SK104A &   11.43463 &   41.95772 &   11.96 &   -172 &   -301 &  &          B016 &   10.18821 &   41.36939 &   21.36 &   -389 &   -406 \\
         B347 &   10.59537 &   41.90760 &   27.32 &   -283 &   -251 &  &          B304 &    9.73726 &   41.17456 &   31.00 &   -414 &   -341 \\
          H26 &   14.86404 &   37.69281 &  284.25 &   -419 &  99999 &  &          B380 &   11.52583 &   42.01472 &   13.23 &   -111 &    -13 \\
       MCGC10 &   16.85966 &   35.78011 &  430.68 &   -300 &   -291 &  &          B033 &   10.35999 &   41.00382 &    5.15 &   -455 &   -439 \\
         B457 &   10.37170 &   42.31032 &   50.04 &   -333 &    -63 &  &          B015 &   10.18757 &   40.99893 &    9.78 &   -467 &   -460 \\
         B236 &   11.28712 &   40.84132 &   37.14 &   -410 &   -411 &  &          B172 &   10.85832 &   41.35884 &    3.49 &   -187 &   -272 \\
          H17 &   10.59870 &   37.24308 &  151.04 &   -233 &  99999 &  &          B129 &   10.70135 &   41.41852 &    5.18 &   -109 &    -75 \\
         B013 &   10.16022 &   41.42328 &   24.28 &   -422 &   -409 &  &          B372 &   11.38908 &   42.00678 &   12.43 &   -226 &   -216 \\
         V031 &   10.30100 &   41.09133 &    8.28 &   -480 &   -433 &  &          B010 &   10.13154 &   41.23956 &   18.91 &   -195 &   -161 \\
         B196 &   10.95239 &   40.71025 &   30.40 &   -330 &   -313 &  &          M086 &   11.36870 &   41.82488 &   10.71 &    -56 &    -32 \\
         B287 &   11.36873 &   41.50128 &   16.90 &   -226 &   -281 &  &          B477 &   11.28471 &   41.66053 &   10.31 &   -123 &   -110 \\
        B106D &   10.97687 &   41.25381 &   11.02 &   -282 &   -312 &  &          B116 &   10.64390 &   41.54760 &   11.98 &   -357 &   -337 \\
         B290 &    8.58725 &   41.47169 &   82.80 &   -439 &   -381 &  &          B122 &   10.66703 &   41.56299 &   11.81 &   -415 &   -437 \\
         B061 &   10.50050 &   41.49328 &   14.79 &   -284 &   -285 &  &          B376 &   11.45161 &   41.71108 &   14.20 &   -125 &   -163 \\
         B045 &   10.42962 &   41.57222 &   20.20 &   -415 &   -425 &  &          B047 &   10.43982 &   41.70107 &   24.69 &   -299 &   -291 \\
         B228 &   11.13842 &   41.69107 &    7.29 &   -435 &   -457 &  &          B229 &   11.14085 &   41.64127 &    7.16 &    -58 &    -31 \\
         B387 &   11.63959 &   40.73711 &   53.49 &     39 &   -297 &  &          G353 &   12.57577 &   42.59559 &   30.39 &   -297 &   -295 \\
         B182 &   10.90280 &   41.13670 &   12.60 &   -350 &   -347 &  &         MCGC9 &   13.93309 &   42.77114 &   66.06 &   -131 &   -147 \\
         B422 &    9.41034 &   41.99995 &   71.71 &   -212 &   -202 &  &          B316 &    9.97329 &   40.69416 &   11.38 &   -388 &   -348 \\
         B220 &   11.08102 &   41.50969 &    7.15 &   -241 &   -247 &  &          B048 &   10.43965 &   41.22517 &    7.34 &   -359 &   -251 \\
        B337D &   12.29681 &   41.12245 &   62.74 &   -285 &   -222 &  &          B002 &   10.01072 &   41.19822 &   21.89 &   -370 &   -338 \\
       SK086A &   11.10863 &   41.58734 &    6.80 &    -76 &    -69 &  &          B402 &   12.15023 &   42.02624 &   28.73 &   -420 &   -488 \\
         B467 &   10.77679 &   42.03033 &   26.23 &   -294 &   -342 &  &          B160 &   10.81224 &   41.02652 &   13.57 &   -383 &   -354 \\
         B177 &   10.87713 &   41.09509 &   13.24 &   -385 &   -403 &  &          B211 &   11.01214 &   41.33464 &    9.55 &   -211 &    -60 \\
         B367 &   11.19654 &   42.09218 &   17.19 &    -66 &   -152 &  &            H7 &    7.97715 &   40.11312 &   63.17 &   -446 &  99999 \\
         B183 &   10.90391 &   41.03398 &   16.47 &   -203 &   -186 &  &          B378 &   11.48848 &   41.89188 &   12.68 &   -224 &   -205 \\
         B375 &   11.43983 &   41.66175 &   14.94 &   -196 &   -209 &  &          B024 &   10.29938 &   41.76369 &   31.87 &   -232 &   -310 \\
         B307 &    9.82689 &   40.54948 &   13.68 &   -477 &   -397 &  &         B115D &   11.11090 &   41.64960 &    6.74 &   -111 &  99999 \\
         B074 &   10.53354 &   41.72269 &   22.33 &   -463 &   -435 &  &          B336 &   10.19837 &   42.14503 &   49.62 &   -651 &   -609 \\
         B151 &   10.78976 &   41.35892 &    1.66 &   -202 &   -330 &  &          B305 &    9.74522 &   40.27559 &   16.62 &   -436 &   -489 \\
         B056 &   10.46316 &   40.96114 &    5.78 &    -32 &   -382 &  &          B032 &   10.33965 &   41.29170 &   13.19 &   -534 &   -516 \\
         B186 &   10.90923 &   41.60681 &    6.72 &   -144 &   -119 &  &          B303 &    9.71053 &   40.45860 &   15.58 &   -500 &   -464 \\
         V133 &   11.29390 &   42.00333 &   12.89 &   -466 &    -58 &  &          B046 &   10.43593 &   41.77450 &   27.59 &   -370 &    -98 \\
        B051D &   10.58568 &   41.07724 &    4.43 &    -31 &   -209 &  &          B468 &   10.80228 &   39.79926 &   60.17 &   -279 &   -278 \\
       SK053A &   10.67003 &   40.29801 &   36.69 &   -408 &   -348 &  &           H20 &   11.46881 &   39.93116 &   77.96 &    -75 &  99999 \\
         B461 &   10.51773 &   42.05738 &   35.59 &   -307 &   -296 &  &          B075 &   10.53681 &   41.33923 &    7.81 &   -146 &   -212 \\
         B351 &   10.65821 &   42.19192 &   36.16 &   -320 &   -325 &  &          B354 &   10.69835 &   42.00712 &   27.81 &   -186 &   -283 \\
         B396 &   11.85482 &   40.36168 &   75.16 &   -644 &   -561 &  &        SK073A &   10.95221 &   41.13006 &   14.60 &   -413 &    -33 \\
          H16 &   10.15756 &   39.75831 &   41.26 &   -410 &  99999 &  &          B093 &   10.59644 &   41.36207 &    6.55 &   -298 &   -447 \\
         B423 &    9.48630 &   40.95987 &   33.22 &   -270 &   -215 &  &          B309 &    9.85262 &   40.24141 &   17.89 &   -417 &   -480 \\
         B266 &   10.76465 &   41.67541 &   13.04 &   -191 &   -161 &  &          B462 &   10.56134 &   42.02681 &   32.99 &   -250 &   -214 \\
         B237 &   11.28852 &   41.37624 &   18.02 &   -102 &    -86 &  &          B078 &   10.55063 &   41.29967 &    5.88 &   -333 &   -260 \\
         B072 &   10.53087 &   41.37990 &    9.51 &   -142 &    -89 &  &         B041D &   10.51969 &   41.27972 &    6.29 &   -372 &   -289 \\
         B398 &   11.99074 &   41.81264 &   29.10 &   -197 &   -227 &  &          B023 &   10.25494 &   41.22938 &   14.12 &   -343 &   -451 \\
         B260 &   10.63829 &   41.52351 &   11.25 &   -189 &   -190 &  &          B209 &   11.01097 &   41.42406 &    6.87 &   -328 &   -460 \\
         B165 &   10.82590 &   41.18186 &    8.22 &   -191 &    -35 &  &          B095 &   10.60748 &   41.09343 &    4.40 &   -155 &   -233 \\
    \hline
\end{tabular}\\
\end{table*} 

\addtocounter{table}{-1}
\begin{table*}
\scriptsize
 \caption{Continued.}
\begin{tabular}{lcccccclcccccc}
  \hline
  \hline
Name & RA & Dec. & $R_{\rm p}$ & New $V_r $ & RBC $V_r$ & &Name & RA & Dec. & $R_{\rm p}$ & New $V_r $ & RBC $V_r$  \\ 
          & (deg) & (deg) & (kpc) & (\kms) & (\kms) &           &         & (deg) & (deg) & (kpc) & (\kms) & (\kms)          \\
 \hline
         B068 &   10.51336 &   40.98062 &    6.12 &   -310 &   -278 &  &          B109 &   10.63401 &   41.17442 &    2.15 &   -571 &   -372 \\
         B232 &   11.16762 &   41.25012 &   17.98 &   -183 &   -182 &  &         B011D &   10.21508 &   40.73504 &    9.05 &   -538 &   -481 \\
     KHM31-74 &   10.22055 &   40.58880 &   13.14 &    -55 &   -576 &  &      B035 &   10.38578 &   41.64238 &   24.34 &    -45 &    -49 \\
         B342 &   10.35046 &   40.61310 &   15.16 &   -350 &   -479 &  &          B075 &   10.53668 &   41.33922 &    7.81 &   -246 &   -212 \\
         B298 &    9.50106 &   40.73223 &   26.17 &    141 &   -539 &  &          B037 &   10.39570 &   41.24859 &    9.69 &   -304 &   -338 \\
         B283 &   11.23083 &   41.28332 &   19.09 &    -69 &    -83 &  &          B199 &   10.95759 &   40.97071 &   20.72 &   -415 &   -396 \\
         B168 &   10.84385 &   41.73489 &   12.97 &   -139 &   -190 &  &          B341 &   10.28813 &   40.59805 &   14.21 &     -2 &   -321 \\  

\\
         PHAT &   11.24690 &   41.91080 &   11.00 &    -81 &  99999 & &    PHAT &   11.12700 &   41.84364 &   10.47 &   -119 &  99999 \\
   \hline
\end{tabular}\\
\end{table*}

\begin{acknowledgements}

{We thank the referee for helpful comments and suggestions.} This work is partially supported by National Key Basic Research Program of China
2014CB845700 and  China Postdoctoral Science Foundation 2014M560843.

This work has made use of data products from the Guoshoujing Telescope (the
Large Sky Area Multi-Object Fibre Spectroscopic Telescope, LAMOST). LAMOST
is a National Major Scientific Project built by the Chinese Academy of
Sciences. Funding for the project has been provided by the National
Development and Reform Commission. LAMOST is operated and managed by the
National Astronomical Observatories, Chinese Academy of Sciences.

Funding for SDSS-III has been provided by the Alfred P. Sloan Foundation, the Participating Institutions, the National Science Foundation, and the U.S. Department of Energy Office of Science. The SDSS-III web site is http://www.sdss3.org/.

SDSS-III is managed by the Astrophysical Research Consortium for the Participating Institutions of the SDSS-III Collaboration including the University of Arizona, the Brazilian Participation Group, Brookhaven National Laboratory, Carnegie Mellon University, University of Florida, the French Participation Group, the German Participation Group, Harvard University, the Instituto de Astrofisica de Canarias, the Michigan State/Notre Dame/JINA Participation Group, Johns Hopkins University, Lawrence Berkeley National Laboratory, Max Planck Institute for Astrophysics, Max Planck Institute for Extraterrestrial Physics, New Mexico State University, New York University, Ohio State University, Pennsylvania State University, University of Portsmouth, Princeton University, the Spanish Participation Group, University of Tokyo, University of Utah, Vanderbilt University, University of Virginia, University of Washington, and Yale University.
\end{acknowledgements}

\bibliographystyle{raa} 
\bibliography{gc}
\label{lastpage}
\end{document}